\documentclass[showpacs,preprintnumbers,aps,prd,eqsecnum]{revtex4}
\usepackage[dvips]{graphicx}
\usepackage{bm,latexsym,amsmath,amssymb,amsfonts}
\usepackage{bm}
\def\H0{H_0}
\def\hatH{H}

\begin{document}

\title{
Bulk inflaton shadows of vacuum gravity
}

\author{Tsutomu Kobayashi}
\email{tsutomu@tap.scphys.kyoto-u.ac.jp}
\author{Takahiro Tanaka} 
\email{tama@scphys.kyoto-u.ac.jp}

\affiliation{
Department of Physics, Kyoto University, Kyoto 606-8502, Japan 
}

\begin{abstract}

We introduce a $(5+m)$-dimensional vacuum description of
five-dimensional bulk inflaton models with exponential potentials
that makes analysis of cosmological perturbations simple and transparent.
We show that various solutions, 
including the power-law inflation model recently discovered by Koyama and
Takahashi, are generated from known 
$(5+m)$-dimensional vacuum solutions of pure gravity. 
We derive master equations for all types of perturbations, 
and each of them becomes a second order differential equation 
for one master variable supplemented by simple boundary 
conditions on the brane. One exception is the case for massive modes
of scalar perturbations. In this case, there are two 
independent degrees of freedom, and in general it is difficult to
disentangle them into two separate sectors. 
\end{abstract}


\preprint{KUNS-1881}

\maketitle

\section{Introduction}

Recent progress in particle physics suggests that
the universe might be a four-dimensional subspace, called a ``brane'',
embedded in a higher dimensional ``bulk'' spacetime.
In this braneworld picture,
ordinary matter fields are supposed to be confined to the brane,
while gravity can propagate in the bulk.
Various kinds of braneworld models have been proposed,
and the cosmological consequences of these models
have been studied (for a review see, e.g., Ref.~\cite{Langlois:2002bb}).
The idea of the braneworld brings new possibilities, in particular
to scenarios of the early universe.

A simple model proposed by
Randall and Sundrum~\cite{Randall:1999ee,Randall:1999vf} is such that
the unperturbed bulk 
is a five-dimensional anti-de Sitter spacetime (warped bulk)
bounded by one brane or two.
A homogeneous and isotropic cosmological solution based on this model has been
explored~\cite{Binetruy:1999ut, Binetruy:1999hy, Mukohyama:1999qx, Ida:1999ui, Kraus:1999it}.
A slow-roll inflation driven by a scalar field confined to the brane
was considered in~\cite{Maartens:1999hf}. 
An empty bulk, however, seems less likely
from the point of view of unified theories, which often 
require various fields in addition to gravity. 
Considering a bulk scalar field, Himemoto {\em et al}~\cite{Hime-san1, Hime-san2, Minamitsuji}
have shown that, interestingly, a bulk scalar field can mimic 
the standard slow-roll inflation on the brane under a certain condition
(see also Ref.~\cite{Kobayashi:2000yh}).

Also in the context of heterotic M theory 
cosmological solutions 
has been studied~\cite{Lukas:1999yn, Reall:1998mv, Lukas:1998qs, Seto:2000fq}.
In the model discussed in Refs.~\cite{Lukas:1998qs, Reall:1998mv},
the scalar field has an exponential potential in the bulk
and the tensions of the two branes are also exponential functions of the 
scalar field. 
In this model the power-law expansion (but not inflation) is realized on the brane.
A single-brane model with such exponential-type potentials
is also interesting, and it
has been investigated for
a static brane case~\cite{Kachru:2000hf, Cvetic:2000pn, Bozza:2001xt}
and a dynamical (cosmological)
case~\cite{Ochiai:2000qf, Feinstein:2001xs, Langlois:2001dy, Charmousis:2001nq}.
Very recently, an inflationary solution was found
in a similar setup
by Koyama and Takahashi~\cite{Koyama:2003yz, Koyama:2003sb},
extending the results of Refs.~\cite{Cvetic:2000pn, Bozza:2001xt, Ochiai:2000qf, Feinstein:2001xs}.
A striking feature of their model is that 
cosmological perturbations can be solved analytically.

In spite of the tremendous efforts by many
authors~\cite{Mukohyama:2000ui, Kodama:2000fa, Langlois:2000ia, Langlois:2000ph, perturbation_scalar, Langlois:2000ns, Gorbunov:2001ge, Kobayashi:2003cn},
it is still an unsolved problem to calculate the evolution 
of perturbations in the braneworld models with infinite extra dimensions. 
This lack of knowledge constrains the predictability of this 
interesting class of models. 
There is an approximate way to estimate 
the density fluctuations evaluated on the brane, in which  
perturbations in the bulk are neglected~\cite{Maartens:1999hf}. 
However, once we take into account perturbations in the bulk,  
generally we cannot avoid solving partial differential 
equations in the bulk 
with discouragingly complicated boundary conditions. 

Only a few cases are known
where perturbation equations can be analytically
solved~\cite{Langlois:2000ns, Gorbunov:2001ge, Kobayashi:2003cn}. 
One of them is the special class of bulk inflaton 
models mentioned above~\cite{Koyama:2003yz, Koyama:2003sb}. 
In this paper we clarify the reason why the perturbation equations 
are soluble in this special case. 
Based on this notion, 
we present a new systematic method to find a wider class of 
background cosmological solutions 
and to analyze perturbations from them.

This paper is organized as follows.
In the next section, we explain our basic ideas of
constructing background solutions and
of analyzing cosmological perturbations.
In Sec.~III we consider a model with a single scalar field in the bulk,
which is the main interest of this work,
and derive an effective theory on the brane.
Then,  in Sec.~IV
we present some examples of exact solutions for the background cosmology
obtained by making use of the ideas explained in Sec.~II.
Section~V deals with cosmological perturbations.
Section~VI is devoted to discussion.

\section{Basic ideas}

\subsection{Solutions in the Randall-Sundrum vacuum braneworld\label{vacuum_sol->braneworld}}

We begin with a model whose action is given by  
\begin{eqnarray}
S = S_g + S_b,
\label{startingaction}
\end{eqnarray}
where
    \begin{eqnarray}
    S_g = \frac{1}{2 \kappa_{D+1}^2} \int
    d^{D+1} X \sqrt{-G} \left( R[G] - 2 \Lambda_{D+1} \right),
    \label{action_bulk}
    \end{eqnarray}
is the action of $(D+1)$-dimensional Einstein gravity 
with a negative cosmological constant $\Lambda_{D+1} = -D(D-1)/2 \ell^2$, 
    \begin{eqnarray}
    S_b = - \int d^D X \sqrt{-g}~ \sigma,
    \label{action_brane}
    \end{eqnarray}
is the action of a vacuum brane with a tension $\sigma$, and $g$ is
the determinant of the induced metric on the brane. 

We assume $Z_2$ symmetry across the brane,
so that the tension of the brane is determined
by the junction condition as
    \begin{eqnarray}
    \H0^2={\kappa^4_{D+1} \sigma^2\over 4(D-1)^2}-{1\over \ell^2},  
    \label{cosmological_const_on_brane2}
    \end{eqnarray}
where $\H0$ is related to the $D$-dimensional 
cosmological constant induced on the brane $\Lambda_b$ by
    \begin{eqnarray}
    \Lambda_b = \frac{1}{2}(D-1)(D-2) \H0^2,
    \label{cosmological_const_on_brane1}
    \end{eqnarray}
and it represents the deviation of $\sigma$ from the fine-tuned Randall-Sundrum value,
$2(D-1)/\ell \kappa^2_{D+1}$.

One of the key ideas in the present paper is to make use of 
the following well-known fact. 
If a metric $ds^2_{(D)}$ is a solution of the $D$-dimensional
vacuum Einstein equations with a cosmological constant $\Lambda_b$,
    \begin{eqnarray}
    ds^2_{(D+1)} = e^{2\omega(z)}
    \left( dz^2 + ds^2_{ (D)} \right),
    \label{D+1metric}
    \end{eqnarray}
is a solution of the $(D+1)$-dimensional model defined by 
Eq.~(\ref{startingaction}), 
and the warp factor $e^{\omega(z)}$ is given by
    \begin{eqnarray}
    e^{\omega(z)} = \frac{\ell \H0}{\sinh (\H0 z)}. 
    \label{warp_factor}
    \end{eqnarray}
(For a Ricci-flat brane, we have $\Lambda_b = 0$. In this case 
the warp factor reduces to $e^{\omega(z)}=\ell/z$.)
Namely, we can construct a $(D+1)$-dimensional solution in the 
Randall-Sundrum braneworld 
from a vacuum solution of the $D$-dimensional Einstein equations.
A well-known example is the five-dimensional black string solution
obtained from 
the four-dimensional Schwarzschild solution~\cite{Chamblin:1999by}.

\subsection{Bulk inflaton models from dimensional reduction}

We explain how to obtain 
an $(n+2)$-dimensional braneworld model with bulk scalar fields
from $(n+2+\sum j_i)[= D+1]$-dimensional spacetime 
by dimensional reduction.
We use $n$ to represent the number of uncompactified spatial dimensions
on the brane, which is three in realistic models. 
Let us consider $(n+2+\sum j_i)$-dimensional spacetime whose
metric is given by 
    \begin{eqnarray}
   ds^2_{(D+1)} =  G_{AB} dX^A dX^B = {\cal G}_{ab}(x) dx^{a} dx^{b}
    + \sum e^{2 \phi_i (x)} d \sigma^2_i,
    \end{eqnarray}
where $d \sigma^2_i$ is the line element of a $j_i$-dimensional 
constant curvature space with the volume ${\cal V}_i$.   
Here indices $a$ and $b$ run from $0$ to $n+1$, and
$\phi_i$ is assumed to depend only on the $(n+2)$-dimensional 
coordinates $x^a$.

Then dimensional reduction to $n+2$ dimensions yields
    \begin{eqnarray}
    S_g^{(n+2)} 
    &=& \frac{1}{2 \kappa_{n+2}^2} \int
    d^{n+2} x \sqrt{-{\cal G}} ~e^{Q} \Bigg[
    R[{\cal G}] - \sum j_i {\cal G}^{ab} \partial_{a} \phi_i \partial_{b} \phi_i
    + {\cal G}^{ab} \partial_{a} Q \partial_{b} Q
    \nonumber\\
    &&\qquad -2 \Lambda_{D+1} + \sum K_i j_i (j_i - 1) e^{-2 \phi_i} \Bigg],
    \end{eqnarray}
where 
\begin{eqnarray}
  Q := \sum j_i \phi_i,
\end{eqnarray}
$\kappa^2_{n+2} := \kappa^2_{D+1} / \prod
{\cal V}_i$ and $K_i$ represents the signature of the 
curvature of the metric $d\sigma^2_i$: $-1$ (open), 
$0$ (flat) or $1$ (closed).   
Making a conformal transformation to the ``Einstein frame'', 
    \begin{equation}
    \tilde{{\cal G}}_{ab} = e^{2Q/n} {\cal G}_{ab}, 
    \label{tildeG}
    \end{equation}
we have
    \begin{eqnarray}
    S_g^{(n+2)}
    &=& \frac{1}{2 \kappa_{n+2}^2} \int
    d^{n+2} x \sqrt{-\tilde{{\cal G}}} \Bigg[
    R[\tilde{{\cal G}}] - \sum j_i \tilde{{\cal G}}^{ab} \partial_{a} \phi_i \partial_{b} \phi_i
    - \frac{1}{n} \tilde{{\cal G}}^{ab} \partial_{a} Q \partial_{b} Q
    \nonumber\\
    &&-2 \Lambda_{D+1}e^{-2Q/n}
    + \sum K_i j_i (j_i - 1) e^{-2 \phi_i - 2Q/n}
    \Bigg]. 
    \label{action_red_bulk}
    \end{eqnarray}
Notice that the kinetic term of each scalar field has an appropriate
signature since $j_i > 0$.
A parallel calculation gives
    \begin{eqnarray}
    S_b^{(n+2)} 
    &=& - \int d^{n+1} x \sqrt{-q}~ \tilde{\sigma} e^{-Q/n},
    \label{action_red_brane}
    \end{eqnarray}
where $\tilde{\sigma} = \sigma \prod {\cal V}_i$, so that
$\kappa^2_{n+2} \tilde \sigma = \kappa^2_{D+1} \sigma$, and
$q$ is the determinant of the induced metric on the brane, $q_{\hat
a\hat b}:=
\tilde{{\cal G}}_{\hat a\hat b}|_{z=z_b}=e^{2Q/n}{\cal G}_{\hat a\hat b}|_{z=z_b}$. 
The hat upon indices represents restriction to 
the subspace parallel to the brane. Hence $\hat a$ and $\hat b$ run 
from $0$ to $n$.
Here $z_b$ represents the location of the brane. 
In this manner, we can derive $(n+2)$-dimensional braneworld 
models with bulk scalar fields
which have exponential-type potentials both in the bulk and on the brane.

\begin{figure}[tb]
  \begin{center}
    \includegraphics[keepaspectratio=true,height=90mm]{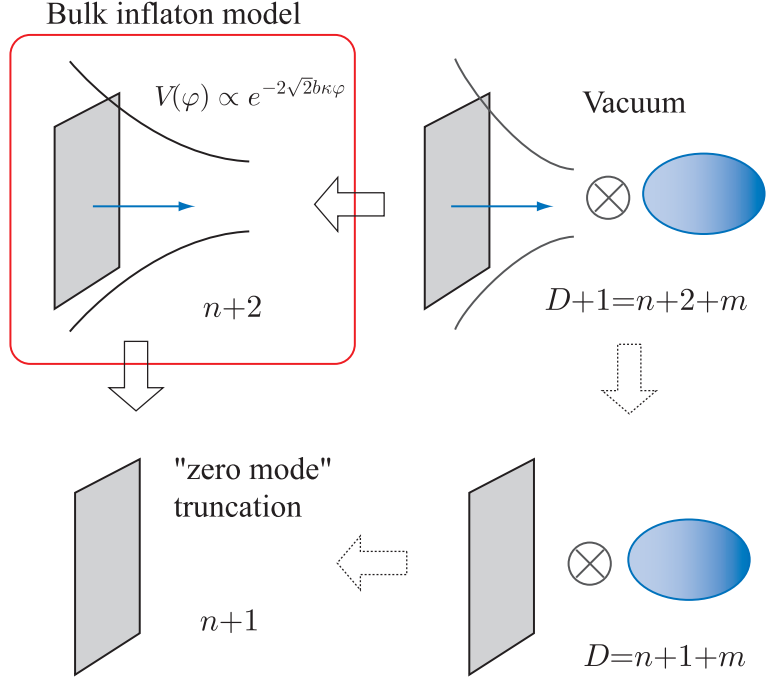}
  \end{center}
  \caption{Schematic of the higher dimensional vacuum description
  of an $(n+2)$-dimensional bulk inflaton model
  and the derivation of the $(n+1)$-dimensional effective theory on the brane.
  The top-left picture represents the bulk inflaton model which
  we are interested in. To analyze cosmological background solutions
  (and ``zero mode'' perturbations),
  we use the $(n+1)$-dimensional
  description shown in the bottom-left corner.
  On the other hand, we make use of the $(D+1)$-dimensional
  description presented in the top-right corner to simplify the
  perturbation analysis.}
  \label{fig:shadow.eps}
\end{figure}

\section{Single scalar field in the bulk \label{4dim-eff-theory}}

From now on, for simplicity, we focus on 
models with a single bulk scalar field. 
The $m$-dimensional space represented by $e^{2\phi} d\sigma^2$
is compactified either on a torus $(K=0)$, on a sphere $(K=1)$, or on a
compact hyperboloid $(K=-1)$. 
Using a canonically normalized 
field $\varphi:=\kappa^{-1}_{n+2} \sqrt{m(m+n)/n}~ \phi$, 
the $(n+2)$-dimensional reduced action is written as
    \begin{eqnarray}
    S^{(n+2)}=\int d^{n+2} x \sqrt{-\tilde {\cal G}}\left\{
    \frac{1}{2\kappa^2_{n+2} }R[\tilde {\cal G}]
    -\frac{1}{2}\tilde {\cal G}^{ab} \partial_a \varphi \partial_b \varphi
    - V(\varphi) \right\} - \int d^{n+1}x \sqrt{-q}~U(\varphi),
    \label{dim_reduced_action}
    \end{eqnarray}
where the potentials are
    \begin{eqnarray}
    && V(\varphi)= -\frac{(m+n)(m+n+1)}{2\kappa_{n+2}^2 \ell^2}
    e^{-2\sqrt{2}b \kappa_{n+2} \varphi}
    - \frac{Km(m-1)}{2\kappa_{n+2}^2} e^{-\sqrt{2}\kappa_{n+2} \varphi/nb},
    \\
    && U(\varphi)= \tilde \sigma e^{-\sqrt{2}b \kappa_{n+2} \varphi},
    \end{eqnarray}
with
    \begin{eqnarray}
    b:=\sqrt{\frac{m}{2n(m+n)}}. 
    \end{eqnarray}

If we assume that 
the metric $ds^2_{(D+1)}$ is given in the form of (\ref{D+1metric}), 
the action can further reduced to 
the $(n+1)$-dimensional effective one on the brane.
We write the $D$-dimensional part of the metric in the form of 
    \begin{eqnarray}
    ds^2_{(D)}=g_{\hat A\hat B}dx^{\hat A}dx^{\hat B}
    =g_{\hat a\hat b}^{(n+1)}(x) dx^{\hat a} dx^{\hat b}
    + e^{2\alpha (t)} d\sigma^2.  
    \label{metric_single_scalar}
    \end{eqnarray}
Then, comparing the coefficient in front of $d\sigma^2$, 
\begin{equation}
  \phi (t,z):= \alpha(t)+\omega(z),  
\end{equation}
follows. 
Also, $(n+2)$-dimensional metric $\tilde{{\cal G}}_{ab}$ is written as  
\begin{equation}
   \tilde {\cal G}_{ab}dx^a dx^b=
   e^{2m\phi/n} {\cal G}_{ab}dx^a dx^b=
   e^{2(m+n)\omega/n}\left(
  q_{\hat a\hat b}dx^{\hat a}dx^{\hat b} +e^{2m\alpha/n}dz^2\right),
\end{equation}
where we have set $e^{\omega(z_b)}=1$.
Substituting the above expression into (\ref{dim_reduced_action}),  
we perform the integration over $z$ to obtain 
\begin{eqnarray}
    S^{(n+1)} 
    = \frac{1}{2\kappa^2_{n+1}} \int d^{n+1}x \sqrt{-q}
    \Bigl\{
    e^{m \alpha /n} R[q] &-& \frac{m(m+n)}{n} e^{m \alpha /n}
    q^{\hat a\hat b}\partial_{\hat a} \alpha \partial_{\hat b} \alpha 
\cr &&
    - 2\Lambda_b e^{-m\alpha /n}
    + Km(m-1)e^{-2\alpha-m\alpha /n} \Bigr\},
    \label{scalar-tensor_theory}
    \end{eqnarray}
where 
\begin{eqnarray*}
  \kappa^2_{n+1}=\kappa^2_{D}/{\cal V}, 
\end{eqnarray*}
with 
\begin{eqnarray*}
    \kappa^2_{D} :=
       \kappa^2_{D+1}\left[ 2\int^{\infty}_{z_b} e^{(m+n)\omega}dz
    \right]^{-1}, 
\end{eqnarray*}
and ${\cal V}$
is the volume of the $m$-dimensional compactified space.

The above reduction to $(n+1)$ dimensions can be done more easily, 
starting with the $(D+1)[=n+2+m]$-dimensional action. 
First we perform the integration over $z$
and obtain a $D$-dimensional effective action,
    \begin{eqnarray}
    S^{(D)} = \frac{1}{2\kappa_D^2}
    \int d^Dx \sqrt{-g}(R[g]-2\Lambda_b),
    \end{eqnarray}
where $\Lambda_b$ is, as before, 
the one defined by Eq.~(\ref{cosmological_const_on_brane1}). 
The obtained effective action is that for $D$-dimensional 
pure gravity with a cosmological constant $\Lambda_b$. 
Compactifying $m$-dimensions further, 
and taking into account 
\begin{equation}
   q_{\hat a\hat b}
  =e^{2m\alpha/n} g_{\hat a\hat b}^{(n+1)}, 
  \label{qab}
\end{equation}
the same expression for the $(n+1)$-dimensional 
effective action~(\ref{scalar-tensor_theory}) can be recovered 
in a parallel way as we did for the reduction from $(D+1)$ dimensions 
to $(n+2)$ dimensions. 
Using $\varphi_{BD}:=e^{m\alpha /n}$ and $\omega_{BD}:=\sqrt{n(m+n)/m}$,
we can rewrite the action into a familiar form:
    \begin{eqnarray}
    S^{(n+1)}= \frac{1}{2\kappa^2_{n+1}} \int d^{n+1}x \sqrt{-q}
    \left\{
    \varphi_{BD}R-\frac{\omega_{BD}}{\varphi_{BD}}(\partial \varphi_{BD})^2
    -2\Lambda_b \varphi_{BD}^{-1}+Km(m-1)\varphi_{BD}^{-(m+2n)/m}
    \right\},
    \label{actionBD}
    \end{eqnarray}
which implies that
the effective theory on the brane is described by
a scalar-tensor theory.
The action~(\ref{scalar-tensor_theory}), or 
equivalently (\ref{actionBD}), describes not only the background 
unperturbed cosmology but also the zero mode perturbation, 
both of which are independent of the extra-dimensional coordinate, $z$, 
apart from the overall factor $e^{2\omega}$.

The induced metric on the brane $q_{ab}$ is the metric in the Jordan frame.
If we use the metric in the Einstein frame, 
    \begin{eqnarray}
    \tilde{q}_{\hat a\hat b} = e^{\frac{2m}{n(n-1)} \alpha} q_{\hat a\hat b}, 
    \label{eff_metric_on_brane}
    \end{eqnarray}
the effective action becomes
    \begin{eqnarray}
    S^{(n+1)} =  \int d^{n+1}x \sqrt{-\tilde q}\left\{
    \frac{1}{2\kappa^2_{n+1}}R[\tilde q]
    -\frac{1}{2}\tilde q^{\hat a\hat b} \partial_{\hat a} \tilde \varphi 
     \partial_{\hat b} 
    \tilde \varphi    - \tilde V (\tilde \varphi ) \right\},
    \label{eff_four_ac}
    \end{eqnarray}
where $\tilde \varphi 
:= \kappa_{n+1}^{-1}\sqrt{m(m+n)/n}~\alpha $ and the potential is
    \begin{eqnarray}
    \kappa_{n+1}^2 \tilde V (\tilde \varphi )
    =\Lambda_b e^{-{2\sqrt2 n\over n-1} b \kappa_{n+1} \tilde \varphi}
    - \frac{Km(m-1)}{2} e^{-{2\sqrt{2}n(m+n-1)\over m(n-1)}b\kappa_{n+1} \tilde \varphi}.
    \label{eff_four_pot}
    \end{eqnarray}
Obviously, the system defined by the above action
is equivalent to Einstein gravity with
a scalar field.
A discussion on this type of potential can be found
in a recent paper by Neupane~\cite{Neupane:2003cs}.

\section{Examples of the background spacetime\label{ex_of_Kasner}}
In this section, 
we give some examples of $D$-dimensional vacuum solutions,
which generate $(D+1)$-dimensional braneworld solutions
by making use of the prescription described in Sec.~\ref{vacuum_sol->braneworld}.
Here we discuss models with a single scalar field 
and investigate their cosmological evolution in detail. 
Generalization to the case of multi scalar fields 
is given in Appendix B.

\subsection{Kasner type solutions\label{ex_of_Kasner_A}}

We first consider the following Kasner-type solution
as an example of the Ricci-flat case, $\Lambda_b=0$,
    \begin{eqnarray*}
    g_{\hat A\hat B}dx^{\hat A}dx^{\hat B} = e^{2 \alpha(\eta)}
    \left[ -d\eta^2 + \gamma_{\mu \nu} dy^{\mu} dy^{\nu} \right]
    +e^{2\beta(\eta)} \delta_{ij} dx^i dx^j,
    \end{eqnarray*}
where $\gamma_{\mu \nu}$ is the metric of $m$-dimensional
constant curvature space, and $i$ and $j$ run from 1 to $n$.
Under the assumption of the above metric form, we solve the vacuum Einstein equations,
    \begin{eqnarray}
    &&e^{2\alpha} R_{\eta}^{~\eta}
    = n \left[ {\beta'}^2 -\alpha' \beta'
    + \beta'' \right]
    +m \alpha'' = 0,
    \label{Kas_eta-eta}
    \\
    &&e^{2\alpha} R_{\mu}^{~\nu}
    = \delta_{\mu}^{~\nu} \left[K(m-1)
    + n\alpha' \beta' +(m-1){\alpha'}^2 + \alpha'' \right]=0,
    \label{Kas_mu-nu}
    \\
    &&e^{2\alpha} R_{i}^{~j}
    = \delta_i{}^j \left[ n{\beta'}^2
    +(m-1)\alpha' \beta' + \beta'' \right]=0,
    \label{Kas_i-j}
    \end{eqnarray}
where prime denotes differentiation with respect to $\eta$.
Eliminating $\alpha''$ and $\alpha'$ from the above three equations, we have
    \begin{eqnarray*}
    \beta'' = \mp \sqrt{
    \frac{n(m+n-1)}{m} {\beta'}^4
    - K (m-1)^2{\beta'}^2}.
    \end{eqnarray*}
We can easily integrate this equation. 
For example, when $K=1$ (compactified on $m$-sphere $S^m$), the solution of 
this equation becomes
    \begin{eqnarray*}
    n \beta' = \frac{\pm (m-1)q}{\sin [(m-1)\eta]},
    \end{eqnarray*}
where
    \begin{eqnarray}
    q := \sqrt{\frac{mn}{m+n-1}},
    \label{def_q}
    \end{eqnarray}
and the integration constant was used to shift the origin of time.
Thus we find
    \begin{eqnarray*}
    e^{\beta}=\left[ \tan
    \left(\frac{m-1}{2}\eta\right)
    \right]^{\pm q/n}.
    \end{eqnarray*}
Substituting this result into Eq.~(\ref{Kas_i-j}), we have
    \begin{eqnarray*}
    \alpha' = \cot[(m-1) \eta]\mp \frac{q}{\sin[(m-1)\eta]}.
    \end{eqnarray*}
This is integrated as 
    \begin{eqnarray*}
    e^{(m-1) \alpha} = \sin [(m-1)\eta]
    \left[ \cot\left( \frac{m-1}{2}\eta \right) \right]^{\pm q}.
    \end{eqnarray*}
The solution for $K=-1$ is easily obtained by replacing
$\sin$, $\tan$ and $\cot,$ in the above expressions by $\sinh, \tanh$
and $\coth$, respectively.
The solution for $K=0$ behaves like
$e^{\beta}\propto \eta^{\pm q/n}$ and $e^{(m-1)\alpha}\propto \eta^{1\mp q}$.

Let us further investigate the cosmology of
the above example. 
Setting $n=3$ and $K=1$, the induced four-dimensional metric becomes
    \begin{equation}
    q_{\hat a\hat b}dx^{\hat a} dx^{\hat b}= 
     e^{2m\alpha /3}\left[-e^{2\alpha}d\eta^2
    +e^{2\beta}\delta_{ij}dx^i dx^j\right],
    \end{equation}
where
    \begin{eqnarray}
    e^{\alpha} & = & \left\{\sin[(m-1)\eta]\right\}^{\frac{1}{m-1}}
    \left\{\cot[(m-1)\eta/2]\right\}^{\frac{q}{m-1}},
    \nonumber
    \\
    e^{\beta} & = & \left\{\tan[(m-1)\eta/2]\right\}^{q/3},
    \label{al_be_K_neq_0}
    \end{eqnarray}
and $q=\sqrt{3m/(m+2)}$~\footnote{
We should remark that
the dynamical solutions in Ref.~\cite{Feinstein:2001xs}
can be obtained if we compactify the $n$-dimensional section
$e^{2\beta}\delta_{ij}dx^idx^j$
and regard the $m$-dimensional section
$e^{2\alpha}\gamma_{\mu\nu}dy^{\mu}dy^{\nu}$ as our 3-space instead.
The flat case ($K=0$) corresponds to the solution in Ref.~\cite{Ochiai:2000qf}.
We should also mention that recently there have been a lot of discussions 
about the solutions with $K=-1$ in an attempt to explain 
accelerated expansion of the universe in the context of M/string
theory
\cite{Neupane:2003cs, Townsend:2003fx, Ohta:2003pu, Roy:2003nd, Emparan:2003gg, Chen:2003ij, Gutperle:2003kc, Chen:2003dc}.
Note, however, that these arguments are not in the braneworld context.
}.
We set $q$ to be positive without any loss of generality 
since the signature of $q$ is flipped by 
a shift of the origin of time, $\eta\to \eta+\pi$.  
Here one remark is in order. 
In the original $(D+1)$-dimensional model 
$m$ represents the number of the compactified dimensions
and therefore is supposed to be an integer.  
However, $m$ is just a number parameterizing the form of 
the scalar field potential 
when we start with the action 
(\ref{dim_reduced_action}) obtained after dimensional reduction.  
We therefore find that $m$ can be any real positive number in this context. 
The positivity needs to be assumed to keep 
the appropriate signature of the kinetic 
term for the scalar field $\phi$, or equivalently to keep the relation 
between $\phi$ and $\varphi$ to be real. (Strictly speaking, the
case with $m<-n=-3$ is also allowed.)
Then, $q$ can be regarded as
a continuous parameter with its range $0< q < \sqrt{3}$.

The cosmological time $\tau$ is related to $\eta$ via
    \begin{eqnarray}
    d\tau = e^{(m+3)\alpha /3} d\eta.
    \label{eta-tau}
    \end{eqnarray}
Recall that $q_{\hat a\hat b}$,  
the metric induced on the brane in the five$[=n+2]$-dimensional 
model (\ref{dim_reduced_action}), is 
related to $g_{\hat a\hat b}$ by Eq.~(\ref{qab}).  
Hence the scale factor associated with the metric $q_{\hat a\hat b}$
is given by $a=e^{m\alpha /3+ \beta}$, 
and therefore the Hubble parameter on the brane
$\hatH :=a^{-1}da/d\tau$ is given by 
$
    \hatH =(m\alpha' /3+ \beta')e^{-m\alpha /3-\alpha}.
$
Substituting the above solution (\ref{al_be_K_neq_0}) 
into these expressions, we obtain 
    \begin{eqnarray*}
    a & = & \left\{\sin[(m-1)\eta]\right\}^{\frac{2q^2}{9(q^2-1)}}
    \left\{\tan[(m-1)\eta/2]\right\}^{\frac{q(q^2-3)}{9(q^2-1)}},
    \\
    \hatH &=&{q\left(q^2-3+2q\cos[(m-1)\eta]\right)\over 3 (q^2 -3)}
    \left\{\sin[(m-1)\eta]\right\}^{-\frac{8 q^2}{9(q^2-1)}}
    \left\{\tan[(m-1)\eta/2]\right\}^{-\frac{q(q^2-9)}{9(q^2-1)}}.
    \end{eqnarray*}

The relation between the coordinate time $\eta$
and the cosmological time $\tau$~(\ref{eta-tau})
is not so obvious, but the asymptotic behavior can be easily studied.
When $\eta\to 0$, we have $\tau \propto \eta^{\frac{q(q+9)}{9(q+1)}} \to 0$ and
    \begin{eqnarray*}
    a & \propto & \eta^{\frac{q(q+3)}{9(q+1)}} \propto \tau^{p_-},
    \\
    \hatH & \propto & \eta^{-\frac{q(q+9)}{9(q+1)}} 
    \propto a^{-\frac{q+9}{q+3}},
    \\
    e^{\alpha} &\propto& \eta^{-\frac{3-q^2}{3(q+1)}} \to +\infty,
    \end{eqnarray*}
where the exponent $p_-$ is defined below in Eq.~(\ref{ppm}).
For $\bar\eta :=\eta-\pi/(m-1)\to 0$,
the cosmological time is (locally) expressed as
$\tau \propto \bar \eta^{-\frac{q(q-9)}{9(q-1)}}$. 
Therefore the range of the proper time $\tau$ is infinite 
for the parameter region $0<q<1$, while 
it is finite for $1<q<\sqrt{3}$. 
In this limit $\bar\eta \to 0$, the scale factor, the Hubble parameter,
and the scalar field behave like
    \begin{eqnarray*}
    a & \propto & \bar \eta^{-\frac{q(q-3)}{9(q-1)}} \propto
    \left\{ {
    \tau^{p_+},~~~~~~~~~~(0<q<1),
    \atop
    (\tau_{end}-\tau)^{p_+},~~~(1<q<\sqrt{3}),
    }\right.
    \\
    \hatH & \propto & \bar \eta^{\frac{q(q-9)}{9(q-1)}} 
    \propto a^{-\frac{q-9}{q-3}},
    \\
    e^{\alpha} &\propto& \bar \eta^{-\frac{3-q^2}{3(1-q)}} \to
    \left\{ {
    +\infty,~~~(0<q<1),
    \atop
    0,~~~~(1<q<\sqrt{3}).
    }\right. 
    \end{eqnarray*}
In the above expressions, we have used
    \begin{eqnarray}
    p_{\pm}:=\frac{m+3}{4m+9\pm\sqrt{3m(m+2)}}.
    \label{ppm}
    \end{eqnarray}
The ranges of $p_+$ and $p_-$ are
$1/(4+\sqrt{3})<p_+<1/3$ and $1/3<p_-<1/(4-\sqrt{3})$.

\begin{figure}[t]
  \begin{center}
    \includegraphics[keepaspectratio=true,height=48mm]{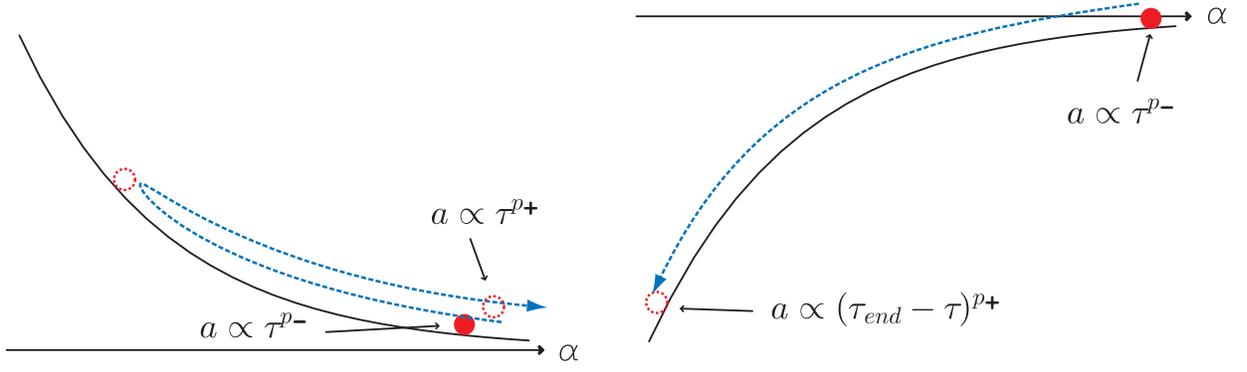}
  \end{center}
  \caption{The motion of the scalar field in the potential.
  The behavior of the scale factor in the Jordan frame
  is also presented.
  The potential is positive for $0<q<1$ (left figure),
  whereas it is negative for $1<q<\sqrt{3}$ (right figure).}
  \label{fig:potential_1.eps}
\end{figure}

The behavior of this solution is
easily understood from the viewpoint of
the four-dimensional effective theory 
described by the action~(\ref{eff_four_ac}),
as was discussed in Ref.~\cite{Emparan:2003gg}.
The potential~(\ref{eff_four_pot}) with $\Lambda_b=0$ and $K=1$
is shown in FIG.~\ref{fig:potential_1.eps}.
For $0<q<1$ ($0<m<1$) the potential is positive,
while for $q>1$ $(m>1)$ the potential is negative.
In the former case, the scalar field $\alpha$
starts at $\alpha = \infty$, climbs up the slope of the potential, 
turns around somewhere, and finally
goes back to $\alpha = \infty$.
In the latter case,
$\alpha$ starts to roll down from $\alpha=\infty$.
The universe expands for a period of time and eventually
it starts to contract. Finally
$\alpha$ falls into the bottomless pit within a finite time,
where the universe ends up with a singularity.

Here we note that 
the above picture based on the four-dimensional effective theory 
describes the dynamics 
in the conformally transformed frame in which 
the metric is given by $\tilde q_{\hat{a} \hat{b}}$, 
whereas we suppose that the ``physical'' metric is
given by the induced metric on the brane $q_{\hat{a} \hat{b}}$. 
In principle, the cosmic expansion law can look very different 
depending on the frame we choose. The dynamics in the 
``physical'' frame therefore can be very different apparently. 
However, the above discussion is still useful since 
the conformal rescaling does not change the 
causal structure of the spacetime.

\begin{figure}[b]
  \begin{center}
    \includegraphics[keepaspectratio=true,height=48mm]{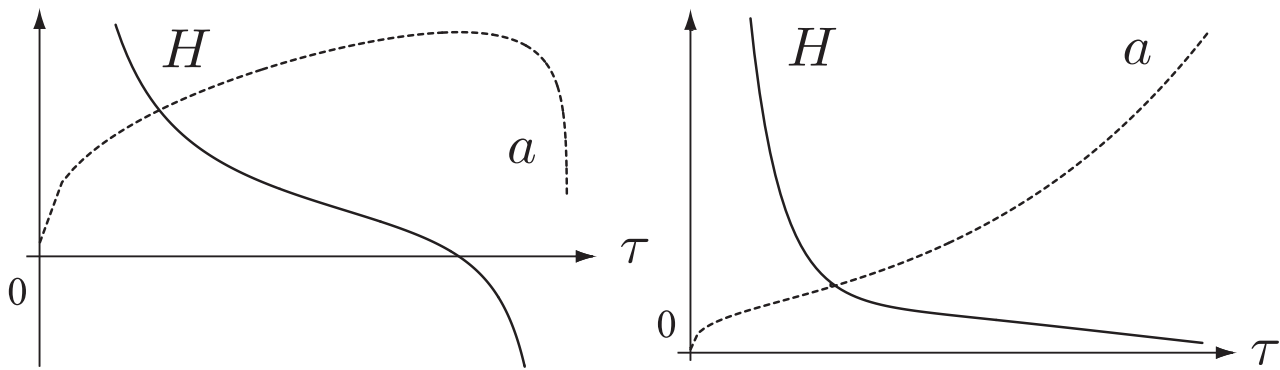}
  \end{center}
  \caption{The sketch of the behavior of the scale factor $a$ (dashed lines)
  and the Hubble parameter $\hatH$ (solid lines) in the Jordan frame
  as functions of the proper time on the brane $\tau$.
  The left figure shows the solution
  of~(\ref{al_be_K_neq_0}) with $q>1$.
  The right figure describes the solution of~(\ref{Kaner_dS_nontrivial})
  with the plus sign in the exponent.}
  \label{fig:Hubble.eps}
\end{figure}

\subsection{Kasner type solutions with a cosmological constant}

The next example is a generalization of the Kasner-type spacetime
including a cosmological constant $\Lambda_b$.
Let us assume that the metric is in the form of
\begin{eqnarray}
g_{\hat A\hat B}dx^{\hat A}dx^{\hat B}
= -dt^2+e^{2\alpha(t)}\delta_{\mu \nu}dy^{\mu}dy^{\nu}
+e^{2\beta(t)}\delta_{ij}dx^idx^j,
\end{eqnarray}
where $i$ and $j$ again run from 1 to $n$,
but here the metric of $m$-dimensional space is 
chosen to be flat ($K=0$) because otherwise the solution 
with $\Lambda_b \neq 0$ is not obtained analytically.
The $D[=n+1+m]$-dimensional vacuum Einstein equations with a cosmological
constant reduce to
    \begin{eqnarray}
    &&m \left(\ddot \alpha + \dot \alpha^2 \right)
    + n \left(\ddot \beta + \dot \beta^2 \right) = (m + n)\H0^2,
    \\
    &&\ddot \alpha + \dot \alpha \left( m\dot \alpha + n\dot \beta \right)
    =(m + n)\H0^2,
    \\
    &&\ddot \beta + \dot \beta \left( m\dot \alpha + n\dot \beta \right)
    =(m + n)\H0^2.
    \end{eqnarray}
From this we obtain two types of solutions (see Appendix B).
One is a trivial solution, namely, $D$-dimensional de Sitter spacetime,
$$
\alpha = \beta = \H0 t.
$$
Of the other type is the following two solutions,
    \begin{eqnarray}
    &&e^{m\alpha+n\beta} = \sinh[(m+n)\H0 t],
    \nonumber\\
    &&e^{\alpha-\beta} =
    \left[ \tanh \left(\frac{m+n}{2} \H0 t \right) \right]
    ^{\pm 1/q},
    \label{Kaner_dS_nontrivial}
    \end{eqnarray}
where $q$ is the one that has been introduced in Eq.~(\ref{def_q}).
The range of the time coordinate $t$ is
$(-\infty, \infty)$ for the former de Sitter solution
and $[0, \infty)$ for the latter non-trivial solutions.
Applying the method discussed in Sec.~\ref{vacuum_sol->braneworld} 
to these solutions, one can construct background solutions 
for a $(D+1)$-dimensional braneworld model.

First we briefly mention the relation to 
the bulk inflaton model
recently proposed by Koyama and Takahashi~\cite{Koyama:2003yz,
Koyama:2003sb}.
Identifying their model parameters $\Delta$ and $\delta$ 
as 
\begin{equation}
\Delta = 4b^2 - 8/3 =-\frac{2(m+4)}{m+3}, 
\qquad 
\delta = \frac{m+4}{4(m+3)}\frac{\ell^2 \H0^2}{1+\ell^2 \H0^2}, 
\end{equation}
we will find that our model is equivalent to theirs. 
The parameter $m$ is supposed to take any positive number.
Thus it follows that $\Delta$ varies 
in the same region, $-8/3 \leq \Delta \leq -2$, 
considered in~\cite{Koyama:2003yz, Koyama:2003sb}. 
The background metric obtained by substituting 
the simplest solution $\alpha =\beta =\H0 t$
is indeed the case discussed in their paper. 

Next we consider the cosmic expansion law. 
We start with the simplest case $\alpha =\beta =\H0 t$. 
The dimensionally reduced metric 
(on the brane) is
    \begin{eqnarray*}
    q_{\hat a\hat b}dx^{\hat a}dx^{\hat b} = e^{2m \H0 t /3} \left(
    -dt^2 + e^{2 \H0 t} \delta_{ij} dx^i dx^j
    \right).
    \end{eqnarray*}
Introducing the cosmological time $\tau$ and the conformal time $\eta$ defined by
    \begin{eqnarray*}
    d\tau = a d\eta = e^{m\H0 t /3} dt,
    \end{eqnarray*}
the scale factor on the brane $a = e^{(m+3)\H0 t /3}$ is
written in terms of $\tau$ or $\eta$ as
    \begin{eqnarray}
    a \propto \tau^{(m+3)/m} \propto \eta^{-(m+3)/3}.
    \label{power-law(K-T)}
    \end{eqnarray}
Since $1 \leq (m+3)/m < \infty$, power-law inflation with any exponent
can be realized.

Furthermore, we have non-trivial solutions~(\ref{Kaner_dS_nontrivial}).
The behavior of the solutions is as follows.
At early times ($t \sim 0$), the scale factor behaves like
$a = e^{m\alpha /3+\beta} \sim t^{1/3}$, and the cosmological time is given by
$d\tau \sim t^{\left(m\pm \sqrt{3m(m+2)} \right)/3(m+3)}dt$
(and so $\tau \to 0$ as $t \to 0$).
Therefore, we have
    \begin{eqnarray}
    a \sim \tau^{p_{\pm}},
    \label{non_tri_start}
    \end{eqnarray}
with $p_{\pm}$ introduced previously, which implies that
the universe is not accelerated at early times.
At late times ($t \to \infty$), we see that $\alpha \to \H0 t$ and
$\beta \to \H0 t$,
and the solution shows power-law expansion as is given in 
Eq.~(\ref{power-law(K-T)}).

\begin{figure}[t]
  \begin{center}
    \includegraphics[keepaspectratio=true,height=48mm]{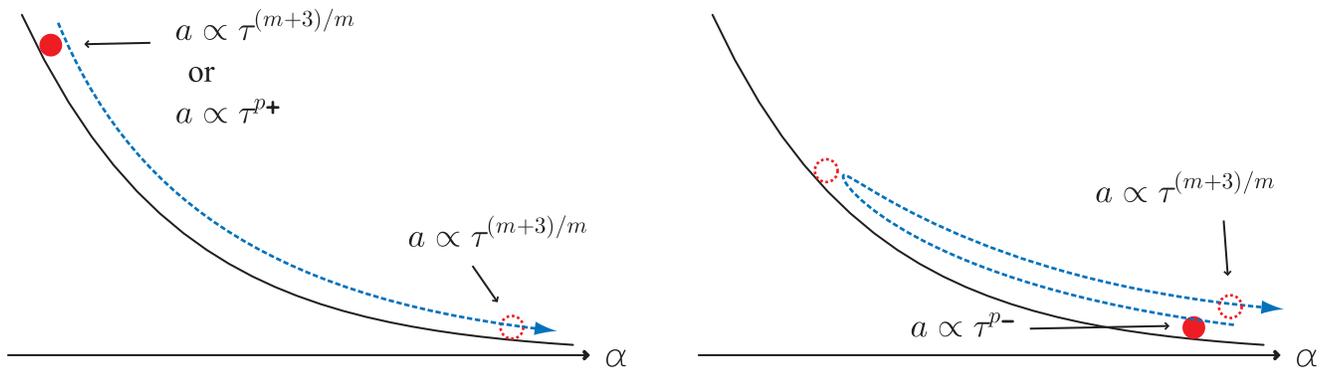}
  \end{center}
  \caption{The motion of the scalar field in the potential
  and the behavior of the scale factor in the Jordan frame.
  The solution found by Koyama and Takahashi~\cite{Koyama:2003yz, Koyama:2003sb}
  is described in the left figure.
  One of the non-trivial solutions with its behavior at the starting point
  $a\propto \tau^{p_+}$ is also shown in the left figure,
  while the other one with the exponent $p_-$ behaves as shown in the right figure.
  }
  \label{fig:potential_2.eps}
\end{figure}

A rather intuitive interpretation of
the behavior of these three solutions
can be made from the four-dimensional point of view again.
The situation is summarized in FIG.~\ref{fig:potential_2.eps}.
This time the potential is always positive. 
For 
one of the non-trivial solutions with the exponent $p_+$
in~(\ref{non_tri_start}),
the scalar field starts to roll down from $\alpha=-\infty$. 
For the other non-trivial solution with the exponent $p_-$,
the field starts to climb up the slope of the potential from $\alpha=+\infty$,
turns around somewhere, and rolls down back to $\alpha=+\infty$.
Suppose that the field $\alpha$ is 
increasing. Let us trace the evolution of $\alpha$ backward in time. 
If the kinetic energy is larger than a certain critical value, 
$\alpha$ will not have a turning point in the past. 
In this case $\alpha$ continues to decrease, reaching $-\infty$. 
This corresponds to the case with the exponent $p_+$.  
If the kinetic energy is lower, $\alpha$ will 
have a turning point. Then, we will have 
$\alpha\to +\infty$ at $t\to -\infty$. This corresponds to the 
case with the exponent $p_-$. 
The case of the power-law inflation~(\ref{power-law(K-T)}) is,  
in fact, the marginal case between these two.
In this case, $\alpha$ does not turn around. 
Therefore the evolution of $\alpha$ is similar to the 
case with $p_+$. 
In any case, 
information about the initial velocity is lost as the universe 
expand. Therefore the late time behavior of the solutions 
is unique, and is given by Eq.~(\ref{power-law(K-T)}).

\section{Cosmological perturbations}

We consider cosmological perturbations in the $(n+2)$-dimensional 
bulk inflaton models defined 
by~(\ref{action_red_bulk}) and~(\ref{action_red_brane}). 
The analysis of perturbations is complicated if we work in the original 
$(n+2)$-dimensional models with a bulk scalar field.
Our $(n+2)$-dimensional system, however,
is equivalent to the $(D+1)[=n+2+m]$-dimensional one 
defined by~(\ref{action_bulk}) and~(\ref{action_brane}). 
We will show that the perturbation analysis becomes 
very simple and transparent in the $(D+1)$-dimensional picture, in which 
we just need to consider pure gravity without any matter fields.

We begin with the following form of background metric:
    \begin{eqnarray}
    G_{AB} d X^A d X^B
    = e^{2 \omega (z)} \left(dz^2 - dt^2 + 
       e^{2 \alpha (t)} \gamma_{\mu \nu} dy^{\mu} dy^{\nu}
    + e^{2 \beta (t)} \delta_{ij} dx^i dx^j\right),
    \end{eqnarray}
where latin and roman indices in the lower case, respectively,  
rum $m$ and $n$-dimensional subspaces, and the warp factor is given by 
$e^{\omega (z)} = \ell \H0 / \sinh (\H0 z)$. We assume that 
$\alpha(t)$ and $\beta(t)$ are chosen so that 
$g_{\hat A\hat B}$ 
be a solution of the $(n+1+m)$-dimensional vacuum Einstein equations
with a cosmological constant, 
    \begin{eqnarray}
    &&m \left(\ddot \alpha + \dot \alpha^2 \right)
    + n \left(\ddot \beta + \dot \beta^2 \right) = (m + n)\H0^2,
    \label{BG_with_K-1}
    \\
    &&\ddot \alpha + \dot \alpha \left( m\dot \alpha + n\dot \beta \right)
    +K(m-1)e^{-2\alpha}
    =(m + n)\H0^2,
    \\
    &&\ddot \beta + \dot \beta \left( m\dot \alpha + n\dot \beta \right)
    =(m + n)\H0^2.
    \label{BG_with_K-3}
    \end{eqnarray}
The background solutions
with $\gamma_{\mu \nu} = \delta_{\mu \nu}~(K=0),~\H0\neq 0$ and
with $K=\pm1,~\H0= 0$ were discussed in the preceding section.
In the following discussions,
we include more general cases with $K=\pm1$ and $\H0\neq 0$.
Although the background solution cannot be obtained in an explicit 
form for such non-flat compactifications with a cosmological constant,
we will find that general properties of perturbations can 
be explored to a great extent.

We write the perturbed metric as
    \begin{eqnarray}
    &&(G_{AB} + \delta G_{AB}) dX^A dX^B 
    \nonumber\\
    &&\qquad =e^{2 \omega} \big\{
    (1 + 2N) dz^2 + 2A dt dz - (1 + 2\Phi) dt^2
    + e^{2 \alpha} (1 + 2S) \gamma_{\mu \nu} dy^{\mu} dy^{\nu}
    \nonumber\\
    &&\qquad\qquad + e^{2 \beta} \left[
    (1 + 2\Psi) \delta_{ij} dx^i dx^j
    + 2E_{ij} dx^i dx^j
    + 2 B_i dx^i dt + 2 C_i dx^i dz \right]
    \big\}.
    \end{eqnarray}
These perturbations are assumed to be homogeneous and 
isotropic in the directions of the $m$-dimensional compactified space 
spanned by the coordinates $y^{\mu}$.
From the assumption of isotropy, mixed components
such as $\delta G_{\mu t}dy^{\mu}dt$ 
and $\delta G_{\mu i}dy^{\mu}dx^i$ are set to zero. 
Concerning the metric perturbations of the compactified space, 
therefore only the overall volume perturbation $S$ is considered.  
After reduction to $(n+2)$-dimensions, $S$ is to be interpreted as the 
scalar field perturbation. 
Here, 
we also assume that the dependence on the $n$-dimensional coordinates
$x^i$ is given by $e^{ik_i x^i}$. 

Metric perturbations are decomposed into
scalar, vector, and tensor components
based on the behavior under the transformation of the $n$-dimensional 
spatial coordinates $x^i$
in the following manner:
    \begin{eqnarray}
    &&B_i = \frac{k_i}{ik} B^S + B^V_i,~~k^i B^V_i = 0,
    \nonumber\\
    &&E_{ij} = \left( - \frac{k_i k_j}{k^2} + \frac{\delta_{ij}}{n} \right)
    E^S + \frac{1}{ik} k_{(i} E_{j)}^V + \frac{1}{2} E_{ij}^T,
    \nonumber\\
    &&k^i E_i^V = k^i E_{ij}^T = 0,~~\delta^{ij} E^T_{ij} = 0.
    \label{decomposition_S-V-T}
    \end{eqnarray}
The quantities with a superscript $S$, $V$ and $T$ represent 
scalar, vector and tensor perturbations, respectively.  
The perturbations $\delta G_{AB}$ obey the linearized Einstein equations
supplemented by boundary conditions at the position of the brane, 
$\delta {\cal K}_A^{~B}\vert_{z=z_b}=0$
where ${\cal K}_{AB}$ is the extrinsic curvature of the brane.
From the $(D+1)$-dimensional point of view
matter sources are absent on the $D$-dimensional brane,
and this makes boundary conditions considerably simple.
Each component of the Einstein equations is presented in Appendix A.

Here we would like to discuss the number of physical degrees of freedom in 
scalar, vector, and tensor perturbations. 
The transverse traceless tensor $E_{ij}^T$ has 
$(n+1)(n-2)/2$ independent components, each of which 
obeys a second order differential equation.
For vector perturbations there are three variables 
$B_i^V$, $E_j^V$ and $C_i^V$. The coordinate transformation 
has one vector mode, and correspondingly there is  
one vector constraint equation. Therefore we have only 
$1[=3-1\times 2]$ vector remaining as a physical mode. 
Since a transverse vector has $(n-1)$ independent components,
we find that 
there are $1\times (n-1)$ degrees of freedom in vector perturbations,
corresponding to the ``graviphoton''.
For scalar perturbations there are 8 variables, and 
the coordinate transformation has 3 independent modes. 
Since there are the same number of constraint equations, 
the number of physical modes is $2[=8-3\times 2]$. 
One of them corresponds to the bulk scalar field and
the other corresponds to the ``graviscalar''.
In total, there are
$1+(n+2)(n-1)/2=2+(n-1)+(n+1)(n-2)/2$ 
physical degrees of freedom. 
The first ``1'' on the left hand side corresponds to the bulk scalar and
the other $(n+2)(n-1)/2$ degrees of freedom to 
those of $(n+2)$-dimensional gravitational waves.

\subsection{Tensor perturbations}

Since tensor perturbations
are gauge-invariant from the beginning,
they are in general easy to analyze. 
The equations 
for tensor perturbations are read from the $\{i,j\}$-component 
of the Einstein equations~(\ref{tens_ij}) as 
    \begin{eqnarray}
    {\cal L} E^T_{ij} = 0,
    \label{eq_for_tensor}
    \end{eqnarray}
where we have defined a differential operator
    \begin{eqnarray*}
    {\cal L}:=\partial^2_t + \left( m \dot \alpha + n \dot \beta \right)\partial_t
    + e^{-2 \beta} k^2 - \partial_z^2- (m+n) (\partial_z\omega) \partial_z.
    \label{Def_L}
    \end{eqnarray*}
The perturbed junction condition 
implies that boundary conditions are Neumann on the brane,
    \begin{eqnarray}
    \left. \partial_z E^T_{ij} \right|_{z=z_b} =0.
    \end{eqnarray}
Since the perturbation equations are manifestly separable,
we write $E^T_{ij}= \chi(t)\psi(z)Y^T_{ij}(x^\ell)$ where
$Y^T_{ij}$ is a transverse, traceless tensor harmonics.
Then $\chi(t)$ and $\psi(z)$ obey
    \begin{eqnarray}
    &&\ddot \chi + \left( m \dot \alpha + n \dot \beta \right) \dot \chi
    + \left( e^{-2 \beta} k^2 + M^2 \right) \chi = 0,
    \label{tensor_t-dep}
    \\
    &&\quad \partial_z^2\psi + (m+n)(\partial_z\omega) \partial_z\psi + M^2 \psi = 0.
    \label{tensor_z-dep}
    \end{eqnarray}
Here $M^2$ is a separation constant and represents the squared Kaluza-Klein mass
for observers on the $D$-dimensional brane.

Now we discuss the mode function in the $z$-direction $\psi(z)$.
Using a canonical variable $\hat \psi := e^{\mu \omega} \psi$,
with
    \begin{eqnarray}
    \mu:=\frac{m+n}{2} \geq \frac{3}{2},
    \label{def_mu}
    \end{eqnarray}
Eq.~(\ref{tensor_z-dep}) is rewritten into a Schr\"{o}dinger-type equation,
    \begin{eqnarray}
    - \partial_z^2\hat\psi + V(z) \hat\psi = M^2 \hat\psi,
    \end{eqnarray}
where the potential is
    \begin{eqnarray}
    V(z) = \mu(\mu +1) \frac{\H0^2}{\sinh^2(\H0 z)}
    + \mu^2 \H0^2-\frac{\kappa_{D+1}^2 \sigma}{2}\delta(z-z_b). 
    \end{eqnarray}
The delta-function term is introduced so that
$\psi$ automatically satisfies the boundary condition $\psi'(z_b)=0$.
The presence of the zero mode, for which $\psi$ is constant in $z$, 
is obvious from Eq.~(\ref{tensor_z-dep}). 
From the asymptotic value of the potential $V(\infty)=\mu^2\H0^2$, 
we can say that
there is a mass gap $\delta M= \mu \H0$ between the zero mode
and the KK continuum.

The $z$-dependence of the massive modes are given
in terms of the associated Legendre functions by
    \begin{eqnarray}
    \psi_M(z) = &
    c^T (\sinh(\H0 z))^{1/2+\mu} & \Biggl[
    Q_{1/2+i\nu}^{-1/2-\mu}(\cosh(\H0 z_b))
    P_{-1/2+i\nu}^{-1/2-\mu}(\cosh(\H0 z)) \cr
   && \qquad - 
    P_{1/2+i\nu}^{-1/2-\mu}(\cosh(\H0 z_b)) 
    Q_{-1/2+i\nu}^{-1/2-\mu}(\cosh(\H0 z))
    \Biggr],
    \label{tensor_KK_Z}
    \end{eqnarray}
where $c^{T}$ is a normalization constant and 
    \begin{eqnarray}
    \nu := \sqrt{\frac{M^2}{\H0^2}-\mu^2}.
    \label{def_nu}
    \end{eqnarray}
These general properties of the mass spectrum and the mode functions in the $z$-direction 
hold irrespective of the specific form of the
background solution $\alpha$ and $\beta$. 

Let us move on to the time dependence of tensor perturbations.
Using the cosmological time on the brane defined by
$\tau =\int e^{m\alpha /n} dt$, Eq.~(\ref{tensor_t-dep}) is rewritten as
    \begin{eqnarray}
    \left[ \frac{d^2}{d\tau^2}
    +\left( n\hatH +\frac{m}{n} \frac{d\alpha}{d\tau}\right)
    \frac{d}{d\tau} + \frac{k^2}{a^2} + e^{-2m\alpha /n} M^2 \right] \chi =0,
    \end{eqnarray}
where one must recall that $a=e^{m\alpha /n+\beta}$ and $\hatH=a^{-1}da/d\tau$.
For the zero mode ($M^2=0$), this reduces to
the equation for the tensor perturbations
in the scalar-tensor theory defined by the action~(\ref{scalar-tensor_theory}).
An apparent difference from Einstein gravity is the presence of the term $(m/n)(d\alpha/d\tau)$.
The Kaluza-Klein mass with respect to observers on the brane
is expressed as
    \begin{eqnarray}
    m^2_{{\rm KK}}(t) = e^{-2m\alpha (t) /n} M^2,
    \end{eqnarray}
and so
$m_{{\rm KK}} = e^{-m\alpha /n}\mu \H0$ for the lightest one,
whereas the Hubble parameter at that time is given by 
    \begin{eqnarray*}
    \hatH = e^{-m\alpha /n} \left(
    \frac{m}{n}\dot \alpha +\dot \beta \right).
    \end{eqnarray*}
For $\alpha=\beta=\H0 t$, this implies $\hatH = 2n^{-1} 
e^{-m\alpha /n}\mu\H0$
and therefore the mass gap and $\hatH$ are
of the same order.
On the other hand, when the background is given by Eq.~(\ref{Kaner_dS_nontrivial}),
we have $\hatH = 2n^{-1} e^{-m\alpha /n}\mu\H0 \coth[(m+n)\H0 t]$
and the mass gap can be very small compared to $\hatH$, 
but only for a short period near $t=0$.

Despite its rather simple form, Eq.~(\ref{tensor_t-dep}) 
cannot be solved analytically in general. 
One exception is the case $\alpha=\beta=\H0 t$ discussed in \cite{Koyama:2003yz}.
In this case, Eq.~(\ref{tensor_t-dep}) reads
    \begin{eqnarray*}
    \ddot \chi + (m+n)\H0 \dot \chi
    + \left( e^{-2\H0 t} k^2 + M^2 \right) \chi = 0,
    \end{eqnarray*}
which, using the conformal time $\eta = -e^{-\H0 t}/\H0$,
can be rewritten as
    \begin{eqnarray}
    \left[ \frac{d^2}{d\eta^2} + \frac{1-m-n}{\eta}\frac{d}{d\eta}
    + k^2 + \frac{M^2}{\H0^2 \eta^2} \right] \chi = 0.
    \end{eqnarray}
This indeed has analytic solutions,
    \begin{eqnarray}
    &&\chi_0 \propto
    (-\eta)^{\mu} H^{(1)}_{\mu}(-k\eta),
    \\
    &&\chi_M \propto
    (-\eta)^{\mu} H^{(1)}_{i\nu}(-k\eta).
    \end{eqnarray}
This is not a surprise 
because the background of the current model is just an
AdS$_{m+2+n}$ bulk with a de Sitter brane.

\subsection{Vector perturbations}
Next we consider vector perturbations. 
From perturbed junction conditions $\delta {\cal K}_i^{~j} |_{z=z_b} = 0$ 
and $\delta {\cal K}_i^{~t}|_{z=z_b} = 0$, we have 
    \begin{eqnarray}
    \left. \partial_z E_i^V \right|_{z=z_b} = 0, ~~\left. {C_i^V} \right|_{z=z_b} = 0,~~
    \left. \partial_z B_i^V \right|_{z=z_b} = 0.
    \label{vec_bc}
    \end{eqnarray}
Under a vector gauge transformation
$x^i \to \bar x^i = x^i + \xi^{iV}$,
metric variables transform as
    \begin{eqnarray}
    \bar E_i^V = E^V_i + k \xi_i^V,
    \qquad
    \bar B_i^V = B^V_i - \dot \xi_i^V,
    \qquad
    \bar C_i^V = C^V_i - \partial_z{\xi_i^V}.
    \end{eqnarray}
Thus we are allowed to set $E_i^V=0$ 
by choosing an appropriate gauge. We expand
the remaining variables by using the transverse vector harmonics $Y^V_i$
as  
    \begin{eqnarray}
    B^V_i = {\cal B} Y^V_i, 
    \qquad
    C^V_i = {\cal C} Y^V_i.  
    \end{eqnarray}
For convenience, we introduce 
    \begin{eqnarray}
    \Omega := k^{-2}
        e^{m\alpha + (n+2)\beta} e^{(m+n)\omega} 
    \left( \partial_z{\cal B} - \dot {\cal C} \right). 
    \label{Def:Omega}
    \end{eqnarray}
Then Eqs.~(\ref{vec_zi}), and~(\ref{vec_ti}) are written as
    \begin{eqnarray}
    {\cal B}
    = e^{-m\alpha - n\beta} e^{-(m+n)\omega} \partial_z\Omega ,
    \\
    {\cal C}
     = e^{-m\alpha + n\beta} e^{-(m+n)\omega} \dot \Omega.  
    \end{eqnarray}
It is easy to see that the remaining third equation is automatically satisfied 
if the above two equations hold. 
Substituting these two into Eq.~(\ref{Def:Omega}), 
we obtain a master equation
    \begin{eqnarray}
    \left[ \ddot \Omega - \left( m \dot \alpha + n \dot \beta \right)
    \dot \Omega + e^{-2\beta} k^2 \Omega \right]
    - \left[ \partial_z^2\Omega - (m+n) (\partial_z\omega) \partial_z\Omega \right] =0.
    \label{master_eq_vec}
    \end{eqnarray}
This equation looks similar to the equation for tensor 
perturbations~(\ref{eq_for_tensor}).
The difference is that the signatures of the terms containing 
first-derivatives such as  
$\dot \Omega$ and $\Omega'$ are reversed.
{}From Eqs.~(\ref{vec_bc}),
the boundary condition for $\Omega$ 
on the brane turns out to be Dirichlet,
    \begin{eqnarray}
    \Omega |_{z=z_b} =0. 
    \end{eqnarray}

Since the master equation~(\ref{master_eq_vec}) is separable,
we write $\Omega(t,z) = \chi(t)\Omega(z)$.
The canonical variable, $\hat \Omega(z) := e^{-\mu \omega} \Omega$,
again obeys a Schr\"{o}dinger-type equation,
    \begin{eqnarray}
    - \partial_z^2{\hat \Omega} + V(z) \hat \Omega = M^2 \hat \Omega,
    \end{eqnarray}
with the potential
    \begin{eqnarray}
    V(z) = \mu (\mu-1) \frac{\H0^2}{\sinh^2(\H0 z)}
    + \mu^2\H0^2.
    \end{eqnarray}
The crucial difference from tensor perturbations is
the absence of the delta-function potential well.
Because of this, there is no zero mode and
only the massive modes with $M^2 \geq V(\infty)= \mu^2 \H0^2$ exist. 
The $z$-dependence of the mode functions are given by
    \begin{eqnarray}
    \Omega_M(z) =
    &c^V(\sinh(\H0 z))^{1/2-\mu}& \Biggl[
    Q_{-1/2+i\nu}^{-1/2+\mu}(\cosh(\H0 z_b)) 
    P_{-1/2+i\nu}^{-1/2+\mu}(\cosh(\H0 z))\cr 
     &&\qquad - P_{-1/2+i\nu}^{-1/2+\mu}(\cosh(\H0 z_b)) 
    Q_{-1/2+i\nu}^{-1/2+\mu}(\cosh(\H0 z))
    \Biggr],
    \label{vector_KK_Z}
    \end{eqnarray}
where $c^V$ is a normalization constant. 
When $\alpha=\beta=\H0 t$, we can find an analytic solution
for the time-dependence of the mode functions,
which, using the conformal time, is given by 
    \begin{eqnarray*}
    \chi_M (\eta) \propto (-\eta)^{-\mu} H^{(1)}_{i\nu}(-k\eta).
    \end{eqnarray*}

\subsection{Scalar perturbations}
\subsubsection{Gauge choice, the boundary condition, and the mode decomposition}
Since scalar perturbations are more complicated, 
we begin with fixing the gauge appropriately in order to 
simplify the perturbed Einstein equations. 
We impose the Gaussian-normal gauge conditions
    \begin{eqnarray}
    N=A=C=0.
    \label{gauge_condi}
    \end{eqnarray}
Different from the case of vector perturbations, these conditions 
do not fix the gauge completely. 
In the case of scalar perturbations we need to take care of 
perturbations of the brane location. 
Here we make use of the remaining gauge degrees of freedom  
to keep the brane location unperturbed at $z=z_b$. 
In the Gaussian-normal gauge, boundary conditions on the brane for all 
remaining variables become Neumann:
    \begin{eqnarray}
   \partial_z \Psi|_{z=z_b} =  \partial_z E|_{z=z_b} =  
   \partial_z S|_{z=z_b} =  \partial_z \Phi|_{z=z_b} =  \partial_z B|_{z=z_b} = 0.
    \end{eqnarray}

Three of eight scalar perturbation equations are the constraint equations,
and the other five are the evolution equations. 
First, let us examine the constraint
equations~(\ref{zi_scalar}),~(\ref{zz_scalar}), and~(\ref{zt_scalar}).
Eq.~(\ref{zz_scalar}) reduces to
$ \partial^2_z(\Phi+n\Psi +mS)+\partial_z \omega  \partial_z(\Phi+n\Psi +mS)=0$.
Taking into account the boundary conditions,
this equation is once integrated to give 
    \begin{eqnarray}
    \partial_z(\Phi+n\Psi +mS) = 0.
    \label{trace_bibun}
    \end{eqnarray}
Eqs.~(\ref{zi_scalar}) and~(\ref{zt_scalar}) reduce to
    \begin{eqnarray}
    &&\partial_z \left[ kB +2 \dot \Phi + 2 \left(m \dot \alpha + n \dot \beta \right) \Phi
    - 2m \dot \alpha S - 2n \dot \beta \Psi \right]=0,
    \label{constraint1}
    \\
    &&\partial_z \left\{ 
      \dot B + \left( m\dot \alpha + n \dot \beta + 2 \dot \beta \right) B
    + 2 e^{-2 \beta} k \left[ \Psi + \left( \frac{1}{n} - 1 \right)E \right]
    \right\} = 0.
    \label{constraint2}
    \end{eqnarray}
With the aid of Eq.~(\ref{trace_bibun}), 
we find that all the perturbation equations are separable. 
Furthermore, the $z$-dependent parts of these equations are
the same as those of tensor perturbations with
the same type of boundary conditions. 
Therefore we can expand all variables 
by using the same mode functions in the $z$-direction 
as those for tensor perturbations:
    \begin{eqnarray}
    &&E=E_0(t)\psi_0(z)+\sum E_M(t)\psi_M(z),
\qquad \Phi=\Phi_0(t)\psi_0(z)+\sum \Phi_M(t)\psi_M(z),
\qquad\cdots,
    \end{eqnarray}
where $\psi_0$ is constant, $\psi_M$ is given by
Eq.~(\ref{tensor_KK_Z}), and $M^2\geq \mu^2 \H0^2$.
Consequently,
Eqs.~(\ref{trace_bibun}),~(\ref{constraint1}), and~(\ref{constraint2})
are automatically satisfied for the zero mode. 
For the massive modes these constraint equations give 
    \begin{eqnarray}
    &&\Phi+n\Psi +mS=0,
    \label{gauge-traceless}
    \\
    &&kB +2 \dot \Phi + 2 \left(m \dot \alpha + n \dot \beta \right) \Phi
    - 2m \dot \alpha S - 2n \dot \beta \Psi =0,
    \label{gauge-transverse1}
    \\
    &&\dot B + \left( m\dot \alpha + n \dot \beta + 2 \dot \beta \right) B
    + 2 e^{-2 \beta} k \left[ \Psi + \left( \frac{1}{n} - 1 \right)E \right]
    =0,  
    \label{gauge-transverse2}
    \end{eqnarray}
where the subscript $M$ was abbreviated.
Note that these three equations
are nothing but components of 
the divergence of the metric perturbations, 
    \begin{eqnarray*}
    \nabla^A \delta G_{zA}
    =
    \nabla^A \delta G_{tA}
    =
    \nabla^A \delta G_{iA}
    =0.
    \end{eqnarray*}
In other words, the transverse traceless conditions
are automatically satisfied if one impose
the Gaussian normal gauge conditions except for the contribution 
coming from the zero mode. 
($\nabla^A \delta G_{zA}$ gives the traceless condition.)
Below we discuss
the KK modes and the zero mode separately.

\subsubsection{KK modes}

By using the constraint equations~(\ref{gauge-traceless})-(\ref{gauge-transverse2}),
the Einstein equations~(\ref{ij__trace}),~(\ref{ij__traceless}),~(\ref{ti_scalar}),~(\ref{tt_scalar}),
and~(\ref{perturbation_munu}) are simplified to give 
    \begin{eqnarray}
    &&{\cal L} \Psi = - \left( \frac{2}{n} - 1 \right) k \dot \beta B
    + 2 \dot \beta \dot \Phi +2(m+n)\H0^2 \Phi,
    \label{scalar_Psi}\\
    &&{\cal L} E = 2 k \dot \beta B,
    \label{scalar_E}\\
    &&{\cal L} S = k \dot \alpha B + 2 \dot \alpha \dot \Phi
    + 2 e^{-2 \alpha} K(m-1)(S - \Phi)+2(m+n)\H0^2 \Phi,
    \label{scalar_S}\\
    &&{\cal L} \Phi
    = - 2 \left( m \ddot \alpha + n \ddot \beta \right) \Phi
    + 2k \dot \beta B + 2m \ddot \alpha S + 2n \ddot \beta \Psi
    +2(m+n)\H0^2 \Phi,
    \label{scalar_Phi}\\
    &&{\cal L} B = - \left( m \ddot \alpha + n \ddot \beta \right) B
    - 4\dot \beta \dot B - 4 \dot \beta^2 B
    -2(m+n)\H0^2B -4 e^{-2 \beta} k \dot \beta \Phi.
    \label{scalar_B}
    \end{eqnarray}
where ${\cal L}$ is defined in Eq.~(\ref{Def_L}).  
Two of them give independent master equations 
for the massive modes, and 
the remaining three equations do not give any new conditions. 
With the aid of the constraint
equations~(\ref{gauge-traceless})-(\ref{gauge-transverse2}), 
Eqs.~(\ref{scalar_Psi}) and (\ref{scalar_Phi}) can be rewritten as 
\begin{eqnarray}
{\cal L} \Psi & = &
  -2(n-2)\dot\beta\left(\dot \alpha -\dot\beta\right)\Psi
  +{4\over n}\dot\beta\dot\Phi
\cr &&\quad
    -2\left\{ \frac{n-2}{n}\dot\beta \left[(m+1)\dot\alpha+n\dot\beta\right]
    -(m+n)H^2 \right\}\Phi,
\label{Psimaster}\\
{\cal L} \Phi & = &
  -4\dot\beta\dot\Phi
    -2\left\{ (m+1)\ddot\alpha+n\ddot\beta +2\dot\beta
    \left[(m+1)\dot\alpha+n\dot\beta\right]-(m+n)H^2\right\}\Phi
\cr &&\quad
   -2n\left[\ddot \alpha -\ddot\beta+2\dot\beta
   \left(\dot\alpha-\dot\beta\right)\right]\Psi.
\label{Phimaster}
\end{eqnarray}
Unfortunately, except for the simplest case (to be discussed later)
we do not know how to disentangle these two equations, 
although there is no problem in solving these equations 
numerically. 
Once we could solve these coupled equations, 
the other variables $S, B, E$ are easily determined 
just by using the constraint equations.

\subsubsection{Zero mode}

To discuss the zero mode,
it is useful to look at the 
cosmological perturbations 
in the corresponding $(n+1)$-dimensional theory 
defined by the action~(\ref{eff_four_ac}). 
In the case of the 
$(n+1)$-dimensional Friedmann universe with a single scalar field,
there is only one physical degree of freedom in scalar perturbations. 
One can derive a second order differential equation for one 
master variable~\cite{Mukhanov:1990me}. 
Back in the braneworld context,
the background metric and its zero-mode perturbations
are also described by the same effective action~(\ref{eff_four_ac}).
Therefore, the analysis of the zero mode 
is no different from the conventional $(n+1)$-dimensional cosmological
perturbation theory. 
Below we will explain this fact more explicitly.

To begin with, we consider $(n+1+m)$-dimensional 
spacetime whose metric is given by
    \begin{eqnarray}
    ds^2 =
    - (1 + 2\Phi) dt^2
    + e^{2 \alpha} (1 + 2S) \gamma_{\mu \nu} dy^{\mu} dy^{\nu}
    + e^{2 \beta} \left[
    (1 + 2\Psi) \delta_{ij} dx^i dx^j
    + 2E_{ij} dx^i dx^j
    + 2 B_i dx^i dt \right],
    \label{4d_4+m_per}
    \end{eqnarray}
where only scalar perturbations are imposed and 
they are again assumed to be homogeneous and 
isotropic with respect to the $m$-dimensional compactified space 
spanned by $y^\mu$. 
Then, the perturbed Einstein equations
$\delta R_{\hat A\hat B} = (m+n)\H0^2 \delta g_{\hat A\hat B}$ 
become identical to 
Eqs.~(\ref{ij__trace}),~(\ref{ij__traceless}),~(\ref{ti_scalar}),~(\ref{tt_scalar}),
and~(\ref{perturbation_munu}) with $N$, $A$, $C$, and the terms
differentiated by $z$ dropped.
Hence, it is manifest that the analysis of 
zero-mode perturbations in our $(n+2+m)$-dimensional spacetime
is equivalent to that of the above system. 

As for perturbations in $(n+2+m)$-dimensional
spacetime, we have already fixed the gauge by imposing 
three gauge conditions (\ref{gauge_condi}). 
However, these gauge conditions do not fix the gauge completely. 
As is manifest from Eqs.~(\ref{App_gaugeT}), gauge transformations 
satisfying $\xi^z=0$ and ${\xi^S}'={\xi^t}'=0$ do not disturb the 
conditions (\ref{gauge_condi}). 
On the other hand, on the $(n+1)$-dimensional side there are two 
scalar gauge transformations 
    \begin{eqnarray*}
    &&t \to \bar t = t+\xi^t
    \\
    &&x^i \to \bar x^i = x^i+k^i \xi^S/ ik. 
    \end{eqnarray*}
The transformation of metric variables under these gauge transformations 
is the same as that obtained by setting $\xi^z=0$ and
${\xi^S}'={\xi^t}'=0$ in the last five equations in~(\ref{App_gaugeT}). 

If we think of the size of compactified dimension $S$ as a scalar field 
in $(n+1)$-dimensional spacetime, the system reduces to a conventional 
$(n+1)$-dimensional model with a scalar field. 
In the conventional cosmological perturbation theory, 
$\Phi$ and $\Psi$ 
in the longitudinal gauge $(B=E=0)$ 
are known to be convenient variables. 
Here one remark is that 
we need to take account of a conformal transformation 
to map the theory to the conventional $(n+1)$-dimensional one, 
    \begin{eqnarray}
    d\tilde s^2 = e^{\frac{2m}{n(n-1)}(\alpha+S)}
    \cdot e^{\frac{2m}{n}(\alpha+S)} ds^2
    = e^{\frac{2m}{n-1}(\alpha+S)} ds^2, 
    \label{4d_conf_metric}
    \end{eqnarray}
which follows from the discussion in Sec.~\ref{4dim-eff-theory}.
Then, the variables corresponding to 
the so-called Sasaki-Mukhanov variables are 
    \begin{eqnarray}
    &&\hat \Phi := \Phi + \frac{m}{n-1} S,
    \\
    &&\hat \Psi := \Psi + \frac{m}{n-1} S, 
    \end{eqnarray}
in the longitudinal gauge.

Eliminating $N$, $A$, $B$, $C$, $E$ and the terms
differentiated by $z$, Eq.~(\ref{ij__traceless}) 
becomes
    \begin{eqnarray}
    \hat \Phi + (n-2) \hat \Psi = 0.
    \label{tashite_zero_4D}
    \end{eqnarray}
Similarly, from Eqs.~(\ref{ti_scalar}),~(\ref{ij__trace}) and 
(\ref{perturbation_munu}), we have
    \begin{eqnarray}
    && (n-1) \dot{\hat{\Psi}}
    + (n-2)\left[m\dot \alpha+(n-1) \dot \beta \right] \hat \Psi
    = -\frac{m+n-1}{n-1} m\dot \alpha S.
    \label{4d_constraint_2}\cr
    && {\cal L}_0 \Psi = - 2(n-1)\dot \beta \dot{\hat{\Psi}}
    +2(m+n)\H0^2 \Phi,\cr
    && {\cal L}_0 S = -2\dot (n-1)\alpha \dot {\hat{\Psi}}
    +2(m+n)\H0^2 \Phi -2K(m-1)e^{-2\alpha}(\Phi-S),
    \end{eqnarray}
where
${\cal L}_0 f := \ddot f + \left(m \dot \alpha +n \dot \beta \right)\dot f + e^{-2\beta}k^2 f$.
Combining all these, we obtain the EOM for $\hat \Psi$:
    \begin{eqnarray}
    {\cal L}_0 \hat \Psi
    -2\left(\dot \beta+ \frac{\ddot \alpha}{\dot \alpha} \right)\dot{\hat \Psi}
    +2(n-2)\left( \ddot \beta - \dot \beta \frac{\ddot \alpha}{\dot \alpha}
    \right)\hat \Psi=0.
    \label{zero_master_t}
    \end{eqnarray}
Using the conformal time $\eta=\int e^{-\beta} dt$, we can rewrite this 
into a more familiar form as
    \begin{eqnarray}
    \hat{\Psi}''+\left[(n-1){\cal H} - 
       2\frac{\alpha''}{\alpha'} \right]
    \hat{\Psi}' + k^2 \hat \Psi
    +2(n-2) \left( {\cal H}' 
     - {\cal H} \frac{\alpha''}{\alpha'} \right) 
     \hat \Psi=0,
    \label{zero_master_eta}
    \end{eqnarray}
where we have defined
    \begin{eqnarray}
    {\cal H} := \left(\ln\tilde a\right)'
    = {1\over n-1}\left[m\alpha + (n-1) \beta\right]' , 
    \end{eqnarray}
and 
$
\tilde a = e^{m\alpha /(n-1) + \beta}
$
is the scale factor in the Einstein frame.

Since there is a mass gap between the zero mode and
the massive modes in general in our models except for a short period in the 
cases of $K\ne 0$, the massive modes would not be excited 
easily. 
Hence, the behavior of the zero mode is especially important.
Since we found that the zero mode can be described by the corresponding 
$(n+1)$-dimensional conventional cosmology, 
it can be easily analyzed in general.

\subsubsection{Exactly solvable case}

Let us consider the simplest background given by $\alpha=\beta=\H0 t$ with $K=0$.
In this special case, scalar perturbations including the KK modes 
are solved exactly.
The most remarkable advantage of our approach may be that
$z$-dependence of the modes can be derived for a general background
as we did in the earlier part of this section. 
The time dependent part, which 
is usually non-trivial especially for the KK modes, is also 
solved easily as shown below when $\alpha=\beta=\H0 t$.

Substituting $\alpha=\beta=\H0 t$ into Eq.~(\ref{Phimaster}),
the equation for $\Phi$ is decoupled first,
    \begin{eqnarray}
    {\cal L} \Phi = -4\H0 \dot \Phi
    -2(m+n+2)\H0^2 \Phi. 
    \end{eqnarray}
By assuming $z$-dependence given in (\ref{tensor_KK_Z}), 
we expand as $\Phi=\Phi_M \psi_M$.
Then, using the conformal time, the above equation is rewritten as
    \begin{eqnarray}
    \left[ \frac{d^2}{d\eta^2} -\frac{m+n+3}{\eta}\frac{d}{d\eta}
    +k^2 + \frac{1}{\eta^2}
    \left( \frac{M^2}{\H0^2}+2(m+n+2) \right) \right] \Phi_M = 0.
    \end{eqnarray}
The solution is given in terms of the Hunkel function by
    \begin{eqnarray}
    \Phi_M=c^S_1 (-\eta)^2 \rho(\eta)
    \end{eqnarray}
with
    \begin{eqnarray*}
    \rho=(-\eta)^{\mu} H^{(1)}_{i \nu}(-k\eta),
    \end{eqnarray*}
where $c^S_1$ is a constant and 
$\mu$ and $\nu$ were defined in Eqs.~(\ref{def_mu}) and~(\ref{def_nu}).
Then, 
substituting this into
Eq.~(\ref{gauge-transverse2}) with the aid of 
Eq.~(\ref{gauge-transverse1}), $B_M$ is immediately obtained as
    \begin{eqnarray}
    B_M= -2c^S_1 k^{-1}\H0
    \left[ (-\eta)^3 \rho' + (2\mu -1)(-\eta)^2 \rho
    \right]. 
    \end{eqnarray}
The result is consistent with the evolution equation for $B$~(\ref{scalar_B}).

Eqs.~(\ref{scalar_E}), (\ref{scalar_S}) and (\ref{scalar_B}) 
 are combined to give a simple equation,
    \begin{eqnarray*}
    {\cal L}[\Psi_M + E_M/n - S_M]=0.
    \end{eqnarray*}
The operator ${\cal L}$ is the one that appeared in 
tensor perturbations, 
and so the mode solutions were already known:
    \begin{eqnarray}
    \Psi_M +E_M /n - S_M=c^S_2\rho,
    \label{Psi+E/n+S}
    \end{eqnarray}
where $c^S_2$ is another constant.
Substituting $\Phi_M$ and $B_M$, 
the constraints~(\ref{gauge-traceless}) and~(\ref{gauge-transverse2})
reduce to two algebraic equations for $\Psi_M$, $E_M$, and $S_M$ as
    \begin{eqnarray}
    n\Psi_M+mS_M&=&-c^S_1(-\eta)^2\rho,
    \label{tbs1}\\
    \Psi_M+(1/n-1)E_M&=&
    -c^S_1(-\eta)^2\rho-c^S_1 k^{-2}
    \left[
    (2\mu-1)\eta \rho'+(\nu^2-3\mu^2+2\mu)\rho
    \right].
    \label{tbs2}
    \end{eqnarray}
Solving Eqs.~(\ref{Psi+E/n+S}), (\ref{tbs1}) and (\ref{tbs2}),
we obtain the expressions for $\Psi_M$, $E_M$, and $S_M$.
Thus, all the metric variables can be analytically 
solved.
Note that one of the above two independent solutions 
were already obtained
in Ref.~\cite{Koyama:2003yz, Koyama:2003sb}. 

The zero-mode solution is also easily obtained.
In this background, the master equation becomes
    \begin{eqnarray}
    {\hat \Psi}''-\frac{m+n-3}{\eta} {\hat \Psi}'
    +k^2\hat \Psi=0,
    \end{eqnarray}
and the solution is
    \begin{eqnarray}
    \hat \Psi = c^S (-\eta)^{\mu-1}H^{(1)}_{\mu-1}(-k\eta).
    \end{eqnarray}
\\

\section{Conclusions}

We have shown that a wide class of braneworld models
with bulk scalar fields can be constructed by dimensional reduction 
from a higher dimensional extension of the Randall-Sundrum model with 
an empty bulk. 
The sizes of compactified dimensions translate into 
scalar fields with exponential potentials both 
in the bulk and on the brane. 
We have mainly concentrated on models with a single scalar field,
which include the power-law inflation solution of Ref.~\cite{Koyama:2003yz, Koyama:2003sb}.

First we have investigated the evolution of five$[=n+2]$-dimensional 
background cosmologies, 
giving an intuitive interpretation based on the four$[=n+1]$-dimensional 
effective description.
Then we have studied cosmological perturbations 
in such braneworld models.
Lifting the models to $(5+m)[=n+2+m]$-dimensions is
a powerful technique for this purpose.
The degrees of freedom of a bulk scalar field in $(n+2)$-dimensions are 
deduced from a purely gravitational theory
in the $(n+2+m)$-dimensional Randall-Sundrum braneworld, 
which consists of a vacuum brane and an empty bulk.
We would like to emphasize that the analysis is greatly simplified
thanks to the absence of matter fields.
From the $(n+2+m)$-dimensional perspective,
we have derived master equations
for all types of perturbations.
We have shown that the mode decomposition is possible 
for all models which are constructed by using 
this dimensional reduction technique. 
Moreover, the dependence in the direction of 
the extra dimension perpendicular to the brane can always be
solved analytically. 

As for scalar perturbations,
there are two physical degrees of freedom for the massive modes
and the equations are not decoupled in general.
For the zero mode, however,
the situation is equivalent to the standard four-dimensional inflation
driven by a single scalar field. Hence, only one degree of freedom is 
physical. Therefore we end up with a single master equation. 
To sum up, our ``embedding and reduction'' approach
enables systematic study of cosmological perturbations
in a class of braneworld models with bulk scalar fields.

In this paper, we have not discussed quantum mechanical aspects. 
In order to evaluate the amplitude of the quantum 
fluctuations, the overall normalization
factor of the perturbations must be determined. 
For this purpose, one needs to 
write down the
perturbed action up to the second order 
written solely in terms of physical degrees of freedom 
as is done in the standard cosmological perturbation theory. 
We would like to return to this problem in a future publication. 

In this paper our investigation is restricted to the parameter region
$m>0$, where $m$ is the number of compactified dimensions. 
If $m>0$, a singularity could exist at $z=\infty$, but it is null.
For $m<-3$, we have the right sign for the kinetic term of the scalar field
and so it is possible to consider such models.
In this parameter region, however, there is a timelike singularity
at $z=0$
and therefore we need a regulator brane to hide it.
This case includes the cosmological solution of heterotic M theory~\cite{Lukas:1998qs}
(which corresponds to $m=-18/5$),
and the analysis of cosmological perturbations in such two-branes models~\cite{Seto:2000fq}
would be also meaningful. 
This issue is also left for future work.

\acknowledgments
The discussions given in Sec.~II and IV were initially developed 
in collaboration with Akihiro Ishibashi and Toby Wiseman. 
We would like to thank them for accepting publishing these results 
as a part of this paper. We also thank
Bruce Bassett for reading our manuscript carefully,
and Hideaki Kudoh for useful discussion and comments.
This work is partly supported by 
Monbukagakusho Grant-in-Aid Nos. 12740154 and 14047212 
and by Inamori foundation. 

\appendix
\section{Perturbed Einstein equations and Gauge transformations}

In this appendix, we write down the components of the
perturbed Einstein equations
    \begin{eqnarray*}
    \delta R_{AB} =
    \nabla_C \nabla_{(A} \delta G_{B)}{}^C
    - \frac{1}{2} \Box \delta G_{AB}
    - \frac{1}{2} \nabla_A \nabla_B \delta G
    = - \frac{1 + m + n}{\ell^2} \delta G_{AB}.
    \end{eqnarray*}
The perturbed quantities are decomposed into
scalar, vector, and tensor components
whose basic definitions are given by Eq.~(\ref{decomposition_S-V-T}).
Note that in the following expressions
no gauge conditions have not been imposed yet.

\begin{itemize}
\item $\{i,j\}$-component\\
    \begin{eqnarray}
    &&\frac{1}{2} \left[
    \ddot h_{ij} + (m \dot \alpha + n \dot \beta) \dot h_{ij}
    - h''_{ij} - (m + n) \omega' h'_{ij}
    - e^{- 2 \beta} \partial^k \partial_k h_{ij}
    + 2 e^{- 2 \beta}  \partial^k \partial_{(i} h_{j)k}
    \right]
    \nonumber\\
    &&- \left[
    \partial_{(i} \dot B_{j)} + (m \dot \alpha + n \dot \beta) \partial_{(i} B_{j)}
    - \partial_{(i} C'_{j)} - (m + n) \omega' \partial_{(i} C_{j)}
    \right] - \delta_{ij} \left(
    \dot \beta ~\partial^k B_k - \omega' \partial^k C_k
    \right)
    \nonumber\\
    &&+ \delta_{ij} \Big\{2(m+n)\H0^2(N- \Phi)+
    2 (1 + m + n) \omega'' N + \omega' N'
    \nonumber\\
    &&- \left[m \dot \alpha + (m + 2n) \dot \beta \right] \omega' A
    - \dot \beta A' - \omega' \dot A
    \nonumber\\
    &&- \omega' \left( \Phi + mS + n \Psi \right)'
    + \dot \beta \left( N - \Phi + mS + n \Psi \right)\dot{}
    \Big\} - e^{-2 \beta} \partial_i \partial_j \left( N + \Phi + mS + n \Psi \right) = 0,
    \label{perturbation_ij}
    \end{eqnarray}
where $h_{ij} = 2 \Psi \delta_{ij} + 2 E_{ij}$ and
the dot (prime) denotes $\partial /\partial t$ ($\partial /\partial z$).
(We use prime to denote differentiation 
with respect to $z$ only in Appendix A.)

\textit{Trace Part}:
    \begin{eqnarray}
    &&\ddot \Psi + ( m \dot \alpha +  n \dot \beta) \dot \Psi + e^{-2 \beta} k^2 \Psi
    - {\Psi}'' - (m + n) \omega' {\Psi}'
    \nonumber\\
    &&- e^{-2 \beta} k^2 \frac{2}{n}
    \left[ \Psi + \left(\frac{1}{n} - 1 \right) E \right]
    - k \left[
    \dot B + \left( m \dot \alpha + 2 n \dot \beta \right) B
    - C' - (m + 2n) \omega' C
    \right] \frac{1}{n}
    \nonumber\\
    &&+ \Big\{2(m+n)\H0^2(N- \Phi)
    \nonumber\\
    &&+ 2 (1 + m + n) \omega'' N + \omega' N'
    - \left[m \dot \alpha + (m + 2n) \dot \beta \right] \omega' A
    - \dot \beta A' - \omega' \dot A
    \nonumber\\
    &&- \omega' \left( \Phi + mS + n \Psi \right)'
    + \dot \beta \left( N - \Phi + mS + n \Psi \right)\dot{}
    \Big\} + e^{- 2 \beta}k^2 \left( N + \Phi + mS + n \Psi \right) \frac{1}{n} = 0.
    \label{ij__trace}
    \end{eqnarray}

\textit{Trace-free Part}:
    \begin{eqnarray}
    &&\ddot E + ( m \dot \alpha +  n \dot \beta) \dot E + e^{-2 \beta} k^2 E
    - {E}'' - (m + n) \omega' {E}'
    +2 e^{-2 \beta} k^2 \left[ \Psi + \left(\frac{1}{n} - 1 \right) E \right]
    \nonumber\\
    &&+ k \left[ \dot B + ( m \dot \alpha + n \dot \beta) B \right]
    - k \left[ {C}' + (m + n) \omega' {C} \right]
    - e^{- 2 \beta}k^2 \left( N + \Phi + mS + n \Psi \right) = 0.
    \label{ij__traceless}
    \end{eqnarray}

\textit{Vector}:
    \begin{eqnarray}
    &&\ddot E^V_i + ( m \dot \alpha +  n \dot \beta) \dot E^V_i
    - {E^V_i}{}'' - (m + n) \omega' {E^V_i}{}'
    \nonumber\\
    &&+ k \left[ \dot B^V_i + ( m \dot \alpha + n \dot \beta) B^V_i \right]
    - k \left[ {C^V_i}{}' + (m + n) \omega' {C^V_i} \right] = 0.
    \label{vec_ij}
    \end{eqnarray}

\textit{Tensor}:
    \begin{eqnarray}
    \left[ \partial_t^2 + ( m \dot \alpha + n \dot \beta) \partial_t
    + e^{-2 \beta} k^2 - \partial_z^2 - (m + n) \omega' \partial_z \right]
    E^T_{ij} = 0.
    \label{tens_ij}
    \end{eqnarray}

\item $\{z,i\}$-component\\
    \begin{eqnarray}
    &&\left( \partial^j h_{ij} \right)' - \left( \partial_i A \right) \dot{}
    - \left( m \dot \alpha + n \dot \beta - 2 \dot \beta \right) \partial_i A
    + \partial^j \partial_i C_j - \partial^j \partial_j C_i
    \nonumber\\
    &&- 2 \partial_i \left( \Phi' + m S' + n \Psi' \right) + 2 (m + n) \omega' \partial_i N
    \nonumber\\
    &&+ e^{2 \beta} \left[
    \ddot C_i - \dot {B_i}' + \left( m \dot \alpha + n \dot \beta + 2 \dot \beta \right)
    \left( \dot C_i - {B_i}' \right) \right] = 0.
    \end{eqnarray}

\textit{Scalar}:
    \begin{eqnarray}
    &&k \bigg[ 2 \Psi'  + 2 \left( \frac{1}{n} - 1 \right) E'
    - \dot A - \left( m \dot \alpha + n \dot \beta - 2 \dot \beta \right) A
    - 2 \left( \Phi' + m S' + n \Psi' \right)
    \nonumber\\
    &&+2 (m + n) \omega' N \bigg] - e^{2 \beta} \left[
    \ddot C - \dot {B}' + \left( m \dot \alpha + n \dot \beta + 2 \dot \beta \right)
    \left( \dot C - B' \right) \right] = 0.
    \label{zi_scalar}
    \end{eqnarray}

\textit{Vector}:
    \begin{eqnarray}
    k {E_i^V}{}' + k^2 C_i^V + e^{2 \beta}
    \left[
    \ddot C_i^V - \dot{B}_i^V{}'
    + \left( m \dot \alpha + n \dot \beta + 2 \dot \beta \right)
    \left( \dot C_i^V - {B_i^V}{}' \right) \right] = 0.
    \label{vec_zi}
    \end{eqnarray}

\item $\{t,i\}$-component\\
    \begin{eqnarray}
    &&\left( \partial^j h_{ij} \right)\dot{} + \left( \partial_i A \right)'
    + ( m + n ) \omega' \partial_i A
    + \partial^j \partial_i B_j - \partial^j \partial_j B_i
    \nonumber\\
    &&+2 \left(m \dot \alpha + n \dot \beta - \dot \beta \right)
    \partial_i \Phi - 2 \partial_i \left( \dot N + m \dot S + n \dot \Psi \right)
    + 2 \partial_i \left[  \dot \beta N + m \left( \dot \beta - \dot \alpha \right) S \right]
    \nonumber\\
    &&+ e^{2 \beta} \left[
    - {B_i}'' + \dot {C_i}' +( m + n ) \omega' \left(\dot C_i - {B_i}'
    \right) \right] = 0.
    \end{eqnarray}

\textit{Scalar}:
    \begin{eqnarray}
    &&k \bigg[ 2 \dot \Psi + 2 \left( \frac{1}{n} - 1 \right) \dot E
    + A' + (m+n) \omega' A + 2 \left(m \dot \alpha + n \dot \beta - \dot \beta \right)
    \Phi - 2 \left( \dot N + m \dot S + n \dot \Psi \right)
    \nonumber\\
    &&+ 2 \dot \beta N + 2m \left( \dot \beta - \dot \alpha \right) S
    \bigg] - e^{2 \beta} \left[
    - {B}'' + \dot {C}' +( m + n ) \omega' \left(\dot C - B'
    \right) \right] = 0.
    \label{ti_scalar}
    \end{eqnarray}

\textit{Vector}:
    \begin{eqnarray}
    k \dot E_i^V + k^2 B_i +
    e^{2 \beta} \left[
    - {B_i^V}{}'' + \dot{C}_i^V{}' +( m + n ) \omega' \left(\dot C_i^V - {B_i^V}{}'
    \right) \right] = 0.
    \label{vec_ti}
    \end{eqnarray}

\item $\{t,t\}$-component\\
    \begin{eqnarray}
    &&k \left( \dot B + 2 \dot \beta B - \omega' C \right)
    - \left( n \Psi + m S \right)\ddot{}
    + \omega' \left(\Phi + n \Psi + m S \right)'
    + \dot A' + (1 + m + n )\omega' \dot A
    \nonumber\\
    &&+ \Phi'' + ( m + n )\omega' \Phi' - e^{- 2 \beta} k^2 \Phi
    + \left( m \dot \alpha + n \dot \beta \right) \dot \Phi
    - 2 n \dot \beta \dot \Psi - 2 m \dot \alpha \dot S
    \nonumber\\
    &&- \ddot N - \omega' N' - 2 (1 + m + n) \omega'' N
    -2(m+n)\H0^2(N- \Phi)= 0.
    \label{tt_scalar}
    \end{eqnarray}

\item $\{z,z\}$-component\\
    \begin{eqnarray}
    &&\ddot N + e^{-2 \beta} k^2 N
    + k \left( C' + \omega' C \right)
    - \left( \Phi + n \Psi + m S \right)''   
    - \omega' \left( \Phi + n \Psi + m S \right)'
    \nonumber\\
    &&- \left( \dot A' + \omega' \dot A \right)
    - \left( m \dot \alpha + n \dot \beta \right)
    \left(A' + \omega' A - \dot N \right)
    + (1 + m + n) \left(\omega' N' + 2 \omega'' N \right) = 0.
    \label{zz_scalar}
    \end{eqnarray}

\item $\{t,z\}$-component\\
    \begin{eqnarray}
    &&k \left( \dot C + 2 \dot \beta C
    + B' \right) + e^{-2 \beta} k^2 A
    + 2 \left( m \dot \alpha + n \dot \beta \right) \Phi'
    + 2 \omega' (m+n) \dot N
    \nonumber\\
    &&- 2 \left( n \dot \Psi' + m \dot S' \right)
    - 2 n \dot \beta \Psi' - 2 m \dot \alpha S'
    - 2(m+n)\H0^2 A= 0.
    \label{zt_scalar}
    \end{eqnarray}

\item $\{\mu,\nu\}$-component\\
    \begin{eqnarray}
    &&\ddot S + \left( 2m \dot \alpha + n \dot \beta \right) \dot S
    - S'' - (2m + n) \omega' S'
    + \left[ e^{-2 \beta} k^2 - 2 e^{-2 \alpha} K (m-1) \right] S
    \nonumber\\
    &&+n \dot \alpha \dot \Psi- \dot \alpha \dot \Phi
    - \omega' \left( \Phi' + n \Psi' \right)
    + 2 e^{-2 \alpha} K (m-1) \Phi
    +k \left( \omega' C - \dot \alpha B \right)
    \nonumber\\
    && + \dot \alpha \dot N + \omega' N' + 2 (1 + m + n) \omega'' N
    - \omega' \dot A - \dot \alpha A'
    - \left[(2m +n) \dot \alpha + n \beta \right] \omega' A
    \nonumber\\
    &&+2(m+n)\H0^2(N- \Phi)= 0.
    \label{perturbation_munu}
    \end{eqnarray}

\end{itemize}

Lastly we summarize the gauge transformations
of the metric variables.
Under a scalar gauge transformation,
    \begin{eqnarray}
    t~\to&&\bar t = t + \xi^t,\nonumber\\
    z~\to&&\bar z = z + \xi^z,\nonumber\\
    x^i~\to&&\bar x^i = x^i + \frac{k^i}{ik}\xi^S,
    \end{eqnarray}
the metric variables transform as
    \begin{eqnarray}
    N~\to&&\bar N = N - {\xi^z}'-\omega' \xi^z,
    \nonumber\\
    A~\to&&\bar A = A + {\xi^t}'-\dot \xi^z,
    \nonumber\\
    C~\to&&\bar C = C + e^{-2\beta}k \xi^z-{\xi^S}'
    \nonumber\\
    B~\to&&\bar B = B - e^{-2\beta}k \xi^t-\dot \xi^S,
    \nonumber\\
    \Phi~\to&&\bar \Phi = \Phi - \dot \xi^t - \omega' \xi^z,
    \nonumber\\
    \Psi~\to&&\bar \Psi = \Psi-\frac{1}{n}k \xi^S
    - \omega' \xi^z - \dot \beta \xi^t,
    \nonumber\\
    E~\to&&\bar E = E + k \xi^S,
    \nonumber\\
    S~\to&&\bar S = S - \omega' \xi^z - \dot \alpha \xi^t.
    \label{App_gaugeT}
    \end{eqnarray}
    
\section{Multi scalar field generalization}
Let us give the generalization of the Kasner type metric
discussed in Sec.~\ref{ex_of_Kasner}.
First we generalize the case without $\Lambda_b$
but including the curvature for one of the spatial sections:
    \begin{eqnarray*}
    g_{MN} dx^M dx^N = e^{2 \alpha(\eta)}
    \left[ -d\eta^2 + \gamma_{\mu \nu} dy^{\mu} dy^{\nu} \right]
    +\sum_{i=1}^{\cal N} e^{2\gamma_i(\eta)} \delta_{MN}^{(i)} dx^M dx^N,
    \end{eqnarray*}
where $\delta_{MN}^{(i)}$ is the metric of a $j_i$-dimensional 
flat space and $\gamma_{\mu\nu}$ is the metric 
of a $\bar m$-dimensional maximally symmetric space. 
As before Here we identified $j_1$ with $n$. 
If we compactify 
$(\bar m+\sum_{i=2}^{\cal N} j_i)$ dimensions leaving $(n+1)$ dimensions, 
the compactified space is divided into ${\cal N}$ sectors having 
different scale factors. 
Then the resulting cosmology after dimensional reduction 
will possess ${\cal N}$ scalar fields.

The set of vacuum Einstein equations becomes
    \begin{eqnarray}
    &&e^{2\alpha} R_{\eta}^{~\eta}
    = \sum j_i {\gamma_i '}^2 -\alpha' u + u' +\bar m \alpha'' = 0,
    \label{gKas_eta-eta}
    \\
    &&e^{2\alpha} R_{\mu}^{~\nu}
    = \delta_{\mu}^{~\nu} \left[K(\bar m-1)
    + \alpha' u +(\bar m-1){\alpha'}^2 + \alpha'' \right]=0,
    \label{gKas_mu-nu}
    \\
    &&e^{2\alpha} R_M^{~N}\delta^{(i)}_{NL}
    = \delta^{(i)}_{ML} \left[ {\gamma_i'} u
    +(\bar m-1)\alpha' \gamma_i' + \gamma_i'' \right]=0,
    \label{gKas_i-j}
    \end{eqnarray}
where we have introduced 
    \begin{eqnarray}
    u := \sum j_i \gamma_i'.
    \label{U_and_alpha}
    \end{eqnarray}
From Eq.~(\ref{gKas_i-j}), we obtain
    \begin{eqnarray}
    u^2+(\bar m-1)\alpha' u + u' = 0.
    \end{eqnarray}
Then, it is easy to see that Eqs.~(\ref{gKas_mu-nu}) 
and (\ref{U_and_alpha}) 
are equivalent to
Eqs.~(\ref{Kas_mu-nu}) and (\ref{Kas_i-j}) in the example of Sec.~\ref{ex_of_Kasner_A}.
by identifying $u$ with  $n\beta'$.
Therefore the solution of Eqs. (\ref{gKas_mu-nu}) and (\ref{U_and_alpha})
for $K=1$ is written as
    \begin{eqnarray}
    u={\pm (\bar m-1)\bar q\over \sin [(\bar m-1)\eta]},~~~
    \alpha'=\cot [(\bar m-1)\eta] \mp {\bar q\over \sin [(\bar m-1)\eta]}.
    \end{eqnarray}
Since these two equations (\ref{gKas_mu-nu}) and (\ref{U_and_alpha}) 
do not have dependence on the number
of dimensions, $\bar q$ has not been fixed yet.
Substituting this solution into Eq. (\ref{gKas_i-j}), we obtain
    \begin{eqnarray*}
    (\bar m-1)\gamma'_i \cot [(\bar m-1)\eta]+\gamma''_i=0,
    \end{eqnarray*}
which is integrated to give
    \begin{eqnarray*}
    \gamma'_i ={(\bar m-1)c_i \over \sin [(\bar m-1)\eta]},
    \end{eqnarray*}
where $c_i$ is an integration constant. 
Substituting this into the remaining equation~(\ref{gKas_eta-eta}) and 
the definition of $u$ (\ref{U_and_alpha}), we find that 
the solution is given by 
    \begin{eqnarray*}
    &&e^{(\bar m-1)\alpha}= \sin [(\bar m-1)\eta]
    \left[\cot \left( \frac{\bar m-1}{2} \eta\right) \right]^{\pm \bar q},
    \\
    &&e^{\gamma_i}=\left[\tan\left({\bar m-1\over 2}\eta\right)\right]^{c_i},
    \end{eqnarray*}
with
    \begin{equation*}
    \sum j_i c_i^2 ={\bar m-\bar q^2\over \bar m-1},~~~ 
    \sum j_i c_i = \pm \bar q.
    \end{equation*}

The next is a generalization of the Kasner-type spacetime
including a cosmological constant $\Lambda_b$. 
Let us assume that the metric is in the form of 
    \begin{eqnarray}
    g_{MN} dx^M dx^N = - dt^2 + \sum_{i=0}^{\cal N}
    e^{2 \gamma_i(t)} \delta_{M N}^{(i)} dx^{a} dx^{b}. 
    \label{metric_Kasner-dS}
    \end{eqnarray}
Here all the spatial sections are taken to be flat ($K_i =0$),
because otherwise an analytic solution with $\Lambda_b \neq 0$ 
cannot be found. 
The Einstein equations 
$
R_{MN}[g] = N \H0^2 g_{MN},
$
with $N :=D-1 = \sum n_i$ reduce to 
    \begin{eqnarray}
    &&\sum j_i \left(\ddot \gamma_i + \dot \gamma_i^2 \right)
    = N \H0^2,
    \label{Kasner_dS_1}
    \\
    &&\ddot \gamma_i + \dot \gamma_i \sum j_i \dot \gamma_i
    = N \H0^2,
    \label{Kasner_dS_2}
    \end{eqnarray}
where dot denotes differentiation with respect to $t$.
These equations admit a trivial solution of 
$D$-dimensional de Sitter spacetime,
    \begin{eqnarray}
    \gamma_i = \H0 t.
    \label{sol1}
    \end{eqnarray}
There is another type of non-trivial solutions.
From Eq.~(\ref{Kasner_dS_2}) we find that $u:=\sum j_i \dot \gamma_i$ obeys
    \begin{eqnarray*}
    \dot u + u^2 = N^2 \H0^2.
    \end{eqnarray*}
The solution for this equation is
    \begin{eqnarray*}
    u=N\H0\coth(N\H0 t).
    \end{eqnarray*}
Substituting this into Eq.~(\ref{Kasner_dS_2}), we obtain
    \begin{eqnarray*}
    \left[ \sinh(N\H0 t) \dot \gamma_i \right]\dot{}
    = N\H0^2 \sinh(N\H0 t).
    \end{eqnarray*}
This can be easily integrated and integration constants are
determined from Eq.~(\ref{Kasner_dS_1}). Then we have
    \begin{eqnarray}
    &&\dot \gamma_i
    = \frac{\H0 \left[ \cosh(N\H0 t) + c_i \right]}{\sinh(N\H0 t)},
    \label{Kasner_dS_sol}
    \\
    &&\sum j_i c_i = 0,~~~\sum j_i c_i^2 = N(N-1).
    \end{eqnarray}
Finally, integrating Eq.~(\ref{Kasner_dS_sol}), we obtain 
    \begin{eqnarray}
    e^{N\gamma_i} = \sinh(N\H0 t) \left[
    \tanh \left( \frac{N\H0}{2} t \right) \right]^{c_i}.
    \label{sol2}
    \end{eqnarray}
Integration constants were removed 
by rescaling the spatial coordinates.
There are only these two types of solutions (\ref{sol1}) and 
(\ref{sol2}) for
Eqs.~(\ref{Kasner_dS_1}) and~(\ref{Kasner_dS_2}).


\end{document}